\documentclass[journal=jctcce,manuscript=article]{achemso}
\usepackage[utf8]{inputenc}
\setkeys{acs}{usetitle = true}
\setkeys{acs}{maxauthors=10,etalmode=truncate}
\usepackage{xspace, amsmath, multirow, setspace, graphicx, subfigure}
\usepackage{bm, dcolumn, epsfig, color, xr, booktabs, longtable, lscape, titlecaps}
\Addlcwords{the of into to in for a an with on and at from by}
\mciteErrorOnUnknownfalse
\usepackage{xr}
\usepackage{cleveref}
\usepackage{amsmath}

\title[]{Spectroscopy, Dynamics and Hydration of S-Nitrosylated
  Myoglobin}

\author{Haydar Taylan Turan$^{1}$ and Markus
  Meuwly$^{1}$} \affiliation[misc]{ $^{1}$Department of Chemistry,
  University of Basel, Klingelbergstrasse 80, Basel, Switzerland}
\email{m.meuwly@unibas.ch}

\keywords{ }

\begin{document}
\doublespace

\maketitle
\thispagestyle{empty}

\begin{abstract}
S-nitrosylation, the covalent addition of NO to the thiol side chain
of cysteine, is an important post-transitional modification that can
alter the function of various proteins. The structural dynamics and
vibrational spectroscopy of S-nitrosylation in the condensed phase is
investigated for the methyl-capped cysteine model system and for
myoglobin. Using conventional point charge and physically more
realistic multipolar force fields for the -SNO group it is found that
the SN- and NO-stretch and the SNO-bend vibrations can be located and
distinguished from the other protein modes for simulations of MbSNO at
50 K. The finding of stable cis- and trans-MbSNO is consistent with
experiments on other proteins as is the observation of buried
-SNO. For MbSNO the observed relocation of the EF loop in the
simulations by $\sim 3$ \AA\/ is consistent with the available X-ray
structure and the conformations adopted by the -SNO label are in good
overall agreement with the X-ray structure. Despite the larger size of
the -SNO group, MbSNO is found to recruit more water molecules within
10 \AA\/ of the modification site than WT Mb due to the stronger
electrostatics. Similarly, when comparing the hydration between the A-
and H-helices they differ by up to 30 \% between WT and MbSNO. This
suggests that local hydration can also be significantly modulated
through nitrosylation.
\end{abstract}

\today

\maketitle

\section{Introduction}
Nitric oxide (NO) is a cell-signaling molecule relevant to function in
the cardiovascular, nervous and immune systems.\cite{sessa1994nitric}
Due to its high biological activity and diffusibility, NO plays an
important role in many biological functions ranging from immune
response, regulation of blood pressure to its function as a
neurotransmitter. \cite{houk2003nitroxyl,kone2003protein}. Furthermore,
its reversible binding to ferrous or ferric heme iron is well
characterized.\cite{cooper1999nitric} One of the well known effects of
NO on muscle tissue is the activation of guanylate cyclase by the
binding of NO to the Heme group, which leads to relaxation of the
smooth muscle.\cite{russwurm2004no} In addition, NO can also play an
important role in post-transitional modifications (PTM) of its target
protein.\cite{astier2012nitric}\\

\noindent
S-Nitrosylation, i.e. the covalent addition of NO to the thiol side
chain of cysteine, is an important PTM that mediates signal
transduction. More than 3000 proteins, which are responsible for a
wide range of cellular functions, have been identified as candidates
for S-nitrosylation under physiological
conditions.\cite{hess2012regulation} S-nitrosylation is reversible,
precisely targeted and regulated by temporal and spatial
arrangements. Specificity of S-Nitrosylation can be governed by
acid-base motifs i.e. electrostatic interactions that affect the
p$K_{\rm a}$ of thiol, and the physiological concentration of
NO.\cite{sun2006s} Further, the relative hydrophobicity of the
surrounding region of the thiol may provide 'hydrophobic motifs' that
affect solvent and co-factor accessibility.\cite{hess2005protein}
Moreover, S-nitrosylation is known to alter the function of a protein
via allosteric
regulation\cite{yasukawa2005s,wu2009aging,chen2008cysteine,barrett2005inhibition}
and can inhibit or promote the formation of sulfide linkage within or
between proteins.\cite{hess2005protein}\\

\noindent
A number of studies have been proposed to characterize the underlying
molecular mechanism of S-Nitrosylation. One of them is based on S-N
decomposition that leads to thiyl radical formation, and thiol
deprotonation as a first step.\cite{zhou2017ab} Alternatively, a
mechanism by which NO reacts directly with reduced thiol to generate a
radical intermediate was proposed\cite{gow1997novel}, or NO$_{2}$
reacts with thiol to produce thiyl radical intermediate, subsequently
attacked by NO radical to form
nitrosothiol.\cite{martinez2013specificity} Finally, co-localization
of nitric oxide synthase (NOS) enzymes with target protein found to be
a determinant of S-nitrosylation under physiological
conditions. \cite{foster2003s}\\

\noindent
Beside the actual mechanism for SNO formation, the detection of the
final product is equally important. A wide range of indirect detecting
techniques have been proposed, considered, and applied. They include
the biotin-switch technique\cite{jaffrey2001protein}, His-tag
switch\cite{camerini2007novel} or fluorogenic
probes\cite{shao2017fluorogenic} which can be utilized to infer
S-nitrosylation. Indirect methods usually break the S-N bond and
capture the signals of the sulfur or nitrogen parts, instead of
addressing the SNO adducts directly. However, such techniques have
limitations and drawbacks in terms of their selectivity and
reproducibility.\cite{wang2011chemical} Indirect methods need
sequential manipulation before final analysis and can be hampered by
decreasing chemoselectivity in each step.\cite{alcock2018chemical}
Also, interference of other species such as NO$_{2}^{-}$ can yield
artifacts in quantitative measurements.\cite{paulsen2013cysteine}\\

\noindent
Direct detection techniques such as gold nanoparticles (AuNP),
phenylmercury, organophosphine or benzenesulfinate probes target the
intact SNO moieties.\citep{alcock2018chemical} One limitation is
incomplete free thiol blockage which can lead to false positive
results with AuNP.\cite{devarie2013direct} Also, depending on the
compound used as a probe, the general applicability can be limited due
to toxicity, as is the case with
phenylmercury.\citep{alcock2018chemical} Further, it was reported that
indirect sunlight can lead to reduction of biotin-HPDP to biotin-SH,
and cause false-positives.\cite{forrester2009detection}\\

\noindent
An alternative direct technique to demonstrate that S-nitrosylation
has occurred is infrared (IR) spectroscopy. IR spectroscopy is a
widely used technique for the structural characterization of
proteins.\cite{koziol2015fast,hamm_zanni_2011} Conformationally
sensitive vibrations within the peptide backbone of the protein are
the amide I (1600 to 1700 cm$^{-1}$), amide II (1510 to 1580
cm$^{-1}$), and amide III (1200 to 1350 cm$^{-1}$) bands which report
on the peptide backbone vibrations. As such, the method provides a
structural fingerprint by which target proteins can be
identified. Similarly, the SNO probe has the SN stretch, SNO bending
and NO stretch modes which provide potentially useful signatures in
the infrared to establish that S-nitrosylation has occurred. The main
advantage of SNO is the chemical uniqueness of the probe. Only
cysteine and methionine residues contain a sulfur atom and SN single
and NO double bonds are absent in unmodified amino acids. On the other
hand the frequency of the NO stretch mode at $\sim 1525$ cm$^{-1}$
overlaps with the amide-II band which makes it potentially challenging
for direct detection.\cite{walsh2007s} There are also experimental
difficulties to measure the IR spectrum of proteins in water. Due to
the overlap between the HOH bending vibration of water and the
conformationally sensitive amide II region at $\sim 1600$ cm$^{-1}$,
the signals can be considerably improved by recording difference
spectra for the $^{14}$NO and $^{15}$NO isotopes\cite{Coyle03}, by
subtracting the water
background\cite{manning2005use,kramer2012toward}, or by subtracting
the spectrum of the WT system.\\

\noindent
Several studies examined the
spectral\cite{mueller1984two,canneva2015conformational,gregori2014vibrational,bartberger2000theory,romeo2001metal,hess2001s}
and
structural\cite{khomyakov2017,timerghazin2008,bignon2018computational}
properties of S-Nitrosothiols in various media. The IR spectra of
methyl thionitrites were studied in argon matrices at 12
K\cite{mueller1984two} and for ethyl
thionitrites\cite{canneva2015conformational} in the gas phase for
both, the cis- and trans-conformers of the two molecules. The NO
stretch frequency was at 1527 cm$^{-1}$ and 1537 cm$^{-1}$ for the
cis-, and at 1548 cm$^{-1}$ and 1559 cm$^{-1}$ for the trans-
conformer for methyl-SNO in the argon matrix and for ethyl-SNO in the
gas phase, respectively. Hence, for both systems the frequency of the
trans- conformer is blue shifted with respect to the cis-orientation
by 21 cm$^{-1}$ and 22 cm$^{-1}$, respectively, whereas depending on
the chemical environment the absolute frequencies vary by some 10
cm$^{-1}$. For S-nitrosoglutathione (GSNO) in solution only the
cis-conformer was reported with a stretch frequency at 1497
cm$^{-1}$.\cite{melvin2019s}\\

\noindent
Computational methods provide a viable alternative to investigate the
relationship between structure, spectroscopy, and
dynamics.\cite{MM.rev:2017} Earlier investigations of the thiol group
included electronic structure studies of small model systems, such as
CH$_3$SNO\cite{timerghazin2008,khomyakov2017} or empirical force
fields based on harmonic potentials using point charge (PC)
models.\cite{han:2008,petrov2013systematic} In the present study, the
spectral and structural properties of wild type (WT) blackfin tuna
myoglobin (Mb) and its S-Nitrosylated analogue are studied by means of
molecular dynamics simulations using Morse potentials and point charge
and multipolar (MTP) force fields for the electrostatics. Two
conformations of S-Nitrosylated myoglobin, cis-MbSNO and trans-MbSNO,
with respect to the C$_\beta $SNO angle are considered. The aim is to
characterize the absorption features in the infrared and the
structural effects induced by S-Nitrosylation of Cys10. Changes in the
local structure are potentially important for modifications in the
function of a protein. Hence, the global and local dynamics of the
protein and its helices and loops, and changes in the solvent shells
around the protein are investigated.\\

\noindent
The present work is structured as follows. First, the force field
parametrization and the atomistic simulations are described. This is
followed by the discussion of calculated IR and power spectra of the
CysNO model system. Then, the IR spectra of WT, cis-MbSNO and
trans-MbSNO are presented and discussed. Finally, structural effects
induced by S-Nitrosylation are discussed and conclusions are drawn.\\

\section{Computational Methods}
\subsection{Molecular Dynamics}
All molecular Dynamics (MD) simulations were performed using the
CHARMM\cite{charmm.prog} software with the
CHARMM36\cite{huang2013charmm36} force field. The equations of motion
were propagated with a leapfrog integrator\cite{hairer2003geometric},
using a time step of $\Delta t = 1$ fs and all bonds involving
hydrogen atoms were constrained using
SHAKE.\cite{ryckaert1977numerical} Non-bonded interactions were
treated with a switch function\cite{steinbach1994new} between 12 and
16 \AA\/ and electrostatic interactions were computed with the
particle mesh Ewald method.\cite{darden1993particle}\\

\noindent
{\it Cys-NO:} For simulating nitrosylated cysteine (Cys-NO), the
molecule was placed in the center of a cubic box (dimensions $25
\times 25 \times 25$ \AA\/$^3$) of TIP3P\cite{jorgensen1983comparison}
water and maintained there with a weak center-of-mass constraint with
a force constant of 1 kcal/mol. Two sets of simulations were carried
out: one using PC electrostatics and the other one with MTP
interactions, see below. First, the systems were heated to 300 K and
equilibrated at this temperature in the $NVT$ ensemble for 500
ps. Production simulations of 10 ns were then performed in the $NVE$
ensemble.\\

\noindent
{\it WT and Nitrosylated Mb:} For the simulations involving wild type
and S-nitrosylated Mb, eight different simulations were set up:
wild-type myoglobin (PDB: 2NRL)\cite{schreiter2007s} at 50 K and 300
K, cis- and trans-S-Nitrosylated myoglobin (S-Nitrosylation at Cys10)
with a PC model for the -SNO moiety at 50 K and 300 K, and cis- and
trans-S-Nitrosylated myoglobin with a multipolar charge model at 50 K
and 300K. Because density for the terminal residue Gly147 was missing
in the 2NRL structure only residues Ala2 to Ser146 were used to set up
the WT and S-nitrosylated protein at position Cys10. With this setup,
simulations for cis-MbSNO and trans-MbSNO starting from the same
initial structure except the dihedral angle $\phi(\rm{C_{\beta}SNO})$
($0^{\circ}$ for cis-MbSNO and $180^{\circ}$ for trans-MbSNO) were
started.\\

\noindent
All systems were solvated in a $80 \times 80 \times 80$ \AA\/$^3$
cubic box of TIP3P\cite{jorgensen1983comparison} water molecules with
buffer regions of 15 \AA\ to the edges of the box. The protein was
weakly constrained to the middle of the simulation box, minimized,
heated to the desired temperature and equilibrated for 500 ps in the
$NVT$ ensemble. Production runs of 10 ns were then performed in the
$NVE$ ensemble.\\

\subsection{Parametrisation of the Force Field}
The CHARMM36 force field was employed to describe the
methyl-terminated Cys residue. The additional parameters required for
Cys-NO were determined from electronic structure calculations.  For
this, the structure of H$_3$C-C$_{3}$H$_{6}$NO$_{2}$S-NO was optimized
at the MP2/aug-cc-pVDZ \cite{head1988mp2,kendall1992electron} level of
theory. All electronic structure calculations were carried out with
Gaussian09.\cite{g09} The nature of the stationary point was verified
by calculating vibrational frequencies. Then the SN bond was scanned
along the bond separation $r$ on a grid between 1.3 and 6 \AA\ whereas
the NO bond was scanned between 0.9 and 5.1 \AA\/, both in increments
of 0.1 \AA\/. The coordinates of all atoms except for the NO moiety
along the SN bond scan, and O for the NO bond scan were frozen.\\

\noindent
Next, the reference MP2 energies were fit to Morse oscillator
functions $V (r ) = D_{\rm e}[1 - exp(-\beta (r-r_{\rm e}))]^{2}$
where $D_{\rm e}$ is the well-depth, $r_{\rm e}$ is the equilibrium
bond distance, and $\beta$ is the parameter that controls the
steepness of the potential. Subsequently, the steepness parameters
$\beta$ are adjusted such as to reproduce experimental frequencies of
the SN and NO stretch vibrations at 520 and 1526 cm$^{-1}$,
respectively.\cite{walsh2007s} The $\beta$ values were 2.171
\AA$^{-1}$ and 1.437 \AA$^{-1}$ for the SN and NO stretch before
fitted to the experimental vibration. For these parametrizations the
energetically more stable cis-conformer was considered. The final
Morse parameters after fitting $\beta$ to the experimentally observed
stretch frequencies are $D_{\rm e} = 202.7$ kcal/mol, $\beta = 1.887$
\AA$^{-1}$, $r_{\rm e} = 1.207$ \AA\ for the NO bond and $D_{\rm e} =
71.1$ kcal/mol, $\beta = 1.987$ \AA$^{-1}$, $r_{\rm e} = 1.843$
\AA\ for the SN bond.\\

\noindent
For the SNO bending, the angle was scanned between $0^\circ$ and
$180^\circ$. For the bending potential, the reference energies were
fitted to $V(\theta) = k_{\theta} (\theta-\theta_{e})^{2}$ where
$k_{\theta}$ is the force-constant and $\theta_{e} = 0$ is the
equilibrium angle. Fitting the force constant to reproduce the
experimentally determined frequency\cite{walsh2007s} of the SNO
bending motion in GSNO at 886 cm$^{-1}$ yields a value of $k_{\theta}
= 132.5$ kcal/mol/rad$^{2}$ and $\theta_e = 115.6^{\circ}$.\\

\noindent
For the electrostatic interactions two different models were
developed. One of them is a standard point charge (PC) model and the
second one uses multipoles (MTPs) on the S, N, and O
atoms.\cite{kramer2012atomic,bereau2013scoring} For this, the charges
of all atoms except for the S, N, and O atom (i.e. the -SNO label)
were those of the CHARMM36 force field\cite{huang2013charmm36} which
were kept constant. This was done in order to maintain the
parametrization of the cysteine residue consistent with the remaining
protein force field. On the other hand, the quality of the ESP fit
will be affected by this strategy, see below. Two models were
considered. The first is a PC model, fitted to the electrostatic
potential (ESP) of the optimized structure of Cys-NO at the
MP2/aug-cc-pVDZ level of theory, see Table \ref{tab:tab1}. Next, these
fitted PCs were frozen at the optimized values and the atomic dipole
and quadrupole moments were fitted to improve the
model.\cite{hedin2016toolkit} All parameters are reported in Table
\ref{tab:tab1}.\\

\noindent
The ESP maps for cis- and trans-CysNO with PCs and MTPs and the
differences between the reference ESP (at the MP2 level) and those
from the PC and MTP models are reported in Figures \ref{sifig:cis_esp}
and \ref{sifig:trans_esp}. The RMSE between reference ESP and the PC
model is 8.36 kcal/mol and 7.44 kcal/mol for cis-CysNO and
trans-CysNO, respectively. This rather larger difference originates
from the fact that the CHARMM36 charges were retained and not allowed
to adjust to the reference ESP in the fit. When including the MTPs on
the -SNO label the RMSE decreased to 4.21 kcal/mol for cis-CysNO and
2.32 kcal/mol for trans-CysNO which is a significant improvement over
a conventional PC model.\\

\begin{table}[h]
\centering
\caption{The PC, dipole and quadrupole parameters generated at
  MP2/aug-cc-pVDZ level which been used in the MTP simulations}
\begin{tabular}{c c |c c c| c c c c c}
\toprule
Parameter&PC [$e$] & \multicolumn{3}{c|}{Dipole [$ea_{0}$]} & \multicolumn{5}{c}{Quadrupole [$ea^{2}_{0}$]} \\
\midrule
&$Q_{00}$&$Q_{10}$&$Q_{1C}$&$Q_{1S}$&$Q_{20}$&$Q_{21C}$&$Q_{21S}$&$Q_{22C}$&$Q_{22S}$ \\
\midrule
S & 0.330 & -0.373 & 0.079 & 0.000 & 0.032 & 0.006 & 0.000 & -0.019 & 0.000 \\
N & -0.203 & 0.000 &0.359&0.128&0.370&0.000&0.000&-0.265&0.006\\
O & -0.127 & 0.104 &-0.135 &0.000&-0.040&0.040&0.000&-0.034&0.000\\
\bottomrule
\label{tab:tab1}
\end{tabular}
\end{table}

\subsection{Infrared Spectrum}
The infrared spectrum was obtained from the Fourier transform of the
dipole moment autocorrelation function. For this, the molecular dipole
moment was calculated from the MD trajectories and the partial
charges. The autocorrelation function where $i$ is the index number of
a snapshot
\begin{equation}
\label{correlation}
C(t) = \sum_{i=1}^{N} \langle\overrightarrow{\mu_{i}}(0) \cdot
\overrightarrow{\mu_{i}}(t) \rangle
\end{equation}
was accumulated over $2^{15}$ time origins to cover 1/3 to 1/2 of the
trajectory. From this, the absorption spectrum is determined according
to
\begin{equation}
\label{ir}
A(\omega) = \omega (1 - e^{-h \omega/(k_{\rm B} T)}) \int C(t)
e^{-i \omega t} dt
\end{equation}
where $T$ is the temperature in Kelvin, $k_{\rm B}$ is the Boltzmann
constant and the integral is determined using a fast Fourier transform
(FFT). IR spectra of WT, cis-MbSNO and trans-MbSNO have been generated
for blocks of 100 ps simulation by correlating over $2^{19}$ time
origins. A total of 100 spectra were generated for each system (total
simulation time of 10 ns) and averaged.\\

\noindent
Additionally power spectra of the NO and SN bonds and the SNO angle
have been calculated from the FFT of the bond length and angle time
series autocorrelation functions. This provides assignments of the
vibrational spectra and allows to detect couplings between
modes.\cite{lammers2007investigating} These power spectra were not
averaged and correlated over $2^{15}$ time origins for the entire
simulation time of 10 ns.\\

\section{Results}
First, the structural dynamics and IR spectroscopy of nitrosylated
cysteine in solution was investigated. This provides a basis for
characterization and interpretation the dynamics and spectroscopy of
cysteine incorporated in a protein, such as myoglobin. Next, the
spectroscopy and conformational dynamics of cis- and trans-MbSNO are
discussed.\\

\subsection{Dynamics and Spectroscopy of CysNO}
\begin{figure}[h!]
\centering \includegraphics[scale=0.70]{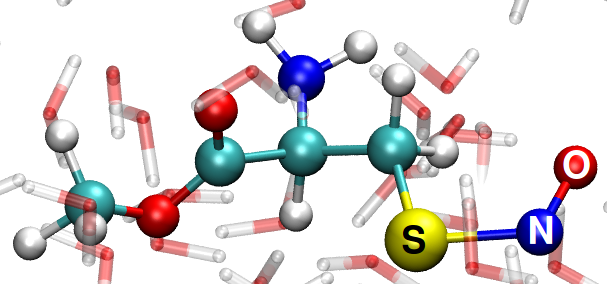}
\caption{The structure of cis-CysNO in aqueous solvent. The CysNO
  ligand is shown as CPK and the TIP3P water molecules are drawn in
  licorice. The color code for the atoms is H (white), C (cyan), O
  (red), N (blue), S (yellow).}
\label{fig:struc.cysno}
\end{figure}

Figure \ref{fig:struc.cysno} shows the structure of cis-CysNO in
aqueous solvent. Quantum chemical calculations at the MP2/aug-cc-pVDZ
level show two minima with $\phi_{\rm CSNO}$ dihedral angles of 0 and
$180^{\circ}$. The energy profile $V(\phi)$ is reported in Figure
\ref{sifig:csno} and shows a stability difference between the global
minimum (cis-Cys-NO) and the trans-conformer ($\phi = 180^\circ$) of
2.5 kcal/mol with the two states separated by a barrier of $\sim 15$
kcal/mol at the MP2 level of theory. Hence, interconversion between
the two states is expected to be slow and will be on the millisecond
time scale according to transition state theory.\\

\begin{figure}[h!]
\centering \includegraphics[scale=0.65]{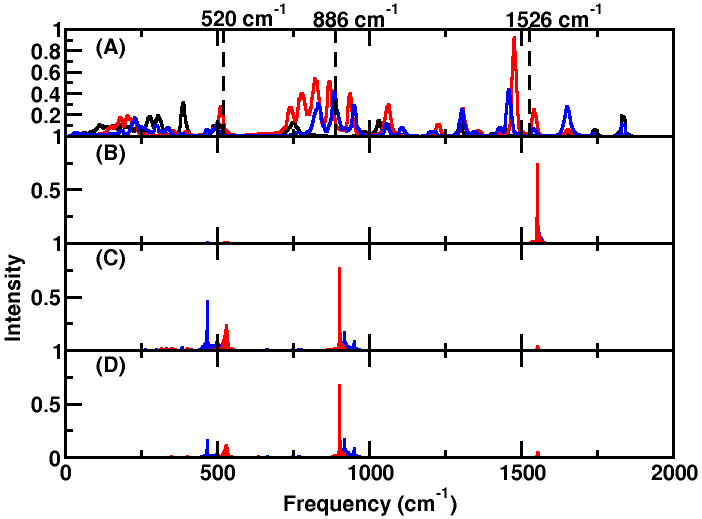}
\caption{Infrared and power spectra for Cys (black) and Cys-NO (blue
  and red) in water. Panel A: calculated IR spectrum of Cys (Black),
  cis-Cys-NO (red) and trans-Cys-NO (blue) from the MD simulations at
  50 K with the MTP model. Panels B to D: the NO, SNO and SN power
  spectra, respectively. The dashed lines at 520 cm$^{-1}$, 886
  cm$^{-1}$ and 1526 cm$^{-1}$ are the experimental values for the SN,
  SNO and NO modes in GSNO in the solid state,
  respectively.\cite{walsh2007s} The dashed lines indicate the
  positions of the experimental values.}
\label{fig:IR_cys}
\end{figure}

\noindent
The IR and power spectra for Cys and CysNO are shown in
Figure~\ref{fig:IR_cys} and compared with the experimental line
positions of the vibrations from Raman spectroscopy of
S-nitrosoglutathione (GSNO) in the solid state\cite{walsh2007s} which
was the reference for the parametrization. The NO stretch peak is
clearly visible in Figure \ref{fig:IR_cys}A at 1540 cm$^{-1}$ for
cis--CysNO and 1538 cm$^{-1}$ for trans--CysNO (red and blue traces,
respectively ) as the power spectrum in Figure \ref{fig:IR_cys}B
confirms. Contrary to that, the SN and SNO peaks are more difficult to
locate in the power spectrum. Considering the power spectra in Figures
\ref{fig:IR_cys}C and D, the SN stretch and SNO bending modes are at
514 cm$^{-1}$ and 871 cm$^{-1}$ for cis-CysNO and 495 cm$^{-1}$ and
881 cm$^{-1}$ for trans-CysNO, respectively, which are close to the
experimental values.  The power spectra (Figures~\ref{fig:IR_cys}B to
D) show that SN and SNO modes are strongly coupled whereas the
coupling to the NO stretch is less pronounced.\\

\noindent
Although the same force field was used for cis- and trans-CysNO, their
spectroscopy for the SN and SNO modes differs. In the power spectra,
the bands of trans-CysNO are red shifted by 29 cm$^{-1}$ for SN and
blue shifted by 14 cm$^{-1}$ for SNO with respect to cis-CysNO, which
should be sufficient to detect both isomers if they are present in
solution. For the NO-stretch the splitting between cis- and
trans-CysNO is 3 cm$^{-1}$ in the gas phase which compares with 21
cm$^{-1}$ for methyl thionitrite (CH$_{3}$SNO) in an Argon
matrix\cite{mueller1984two} whereas in solution, the present
simulations only find an insignificant splitting. Consistent with
experiment, the simulations also find that the trans-conformer absorbs
at higher frequency. The different magnitude of the shift in the gas
phase may be due to both, the different environment (argon in the
experiments) and the somewhat different chemical environment of CH$_3$
versus Cysteine. Further, the SN stretch is at 376 cm$^{-1}$ for
cis-CH$_{3}$SNO and 371 cm$^{-1}$ for trans-CH$_{3}$SNO which amounts
to a red shift by 5 cm$^{-1}$.\cite{mueller1984two} This shift was
more pronounced in the present simulations with 29 cm$^{-1}$. \\

\subsection{The Structural Dynamics and Spectroscopy of Mb-SNO}
\begin{figure}[H]
\centering \includegraphics[scale=0.45]{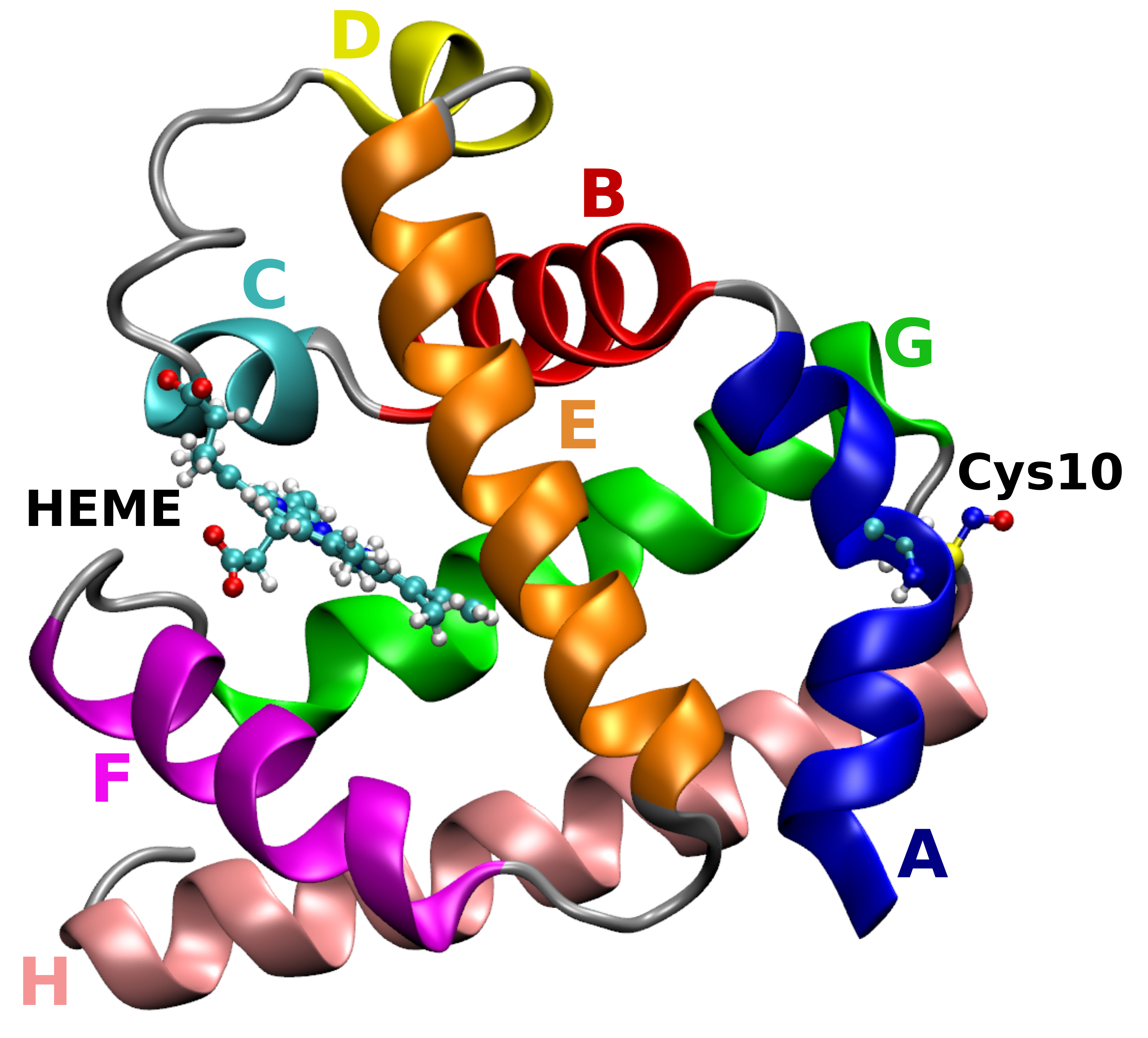}
\caption{Structure of trans-MbSNO (Cartoon representation) with
  helices A to H in different color together with corresponding
  labels. The heme group and the S-Nitrosylated Cys10 residue are
  represented by CPK. Cys10 is part of the A-helix of Mb.}
\label{fig:mbstruc}
\end{figure}

Next, the dynamics and spectroscopy of wild-type (WT) blackfin tuna
myoglobin (PDB: 2NRL) \cite{schreiter2007s} and its S-nitrosylated
variant\cite{schreiter2007s} at Cys10 (PDB: 2NRM) was considered, see
Figure \ref{fig:mbstruc}. For the S-nitrosylated cysteine in blackfin
tuna myoglobin two different cis-conformers were
reported.\cite{schreiter2007s} The major and minor cis- conformers had
$\phi(\rm{NC_{\alpha}C_{\beta}S})$ dihedral angles of $-62^{\circ}$,
$55^{\circ}$ and $\phi(\rm{C_{\alpha}C_{\beta}SN})$ of $172^{\circ}$,
$-72^{\circ}$ in the X-Ray structures, respectively. In the
simulations at 300 K with both, PC and MTP models, the structure with
$\phi(\rm{NC_{\alpha}C_{\beta}S}) \sim -62^{\circ}$ was present
throughout the 10 ns simulation (see Figure
\ref{sifig:dihedral_300}B). With respect to the dihedral
$\phi(\rm{C_{\alpha}C_{\beta}SN})$, the simulation using MTPs sampled
only the state with $\sim 160^{\circ}$ which is close to the major
cis- conformer value whereas the simulation with PCs sampled both
states with $\sim -85^{\circ}$ which is close to the minor cis-
conformer value and $\sim 160^{\circ}$, see Figure
\ref{sifig:dihedral_300}A. The first, shorter lived state with
$\phi(\rm{NC_{\alpha}C_{\beta}S}) \sim -85^{\circ}$ using PCs was
sampled for the first 2 ns after which
$\phi(\rm{NC_{\alpha}C_{\beta}S})$ switched to $\sim 160^{\circ}$ and
remained there for the rest of the simulation (see
Figure~\ref{sifig:dihedral_300}B). These results indicate that the
simulations are able to account for the major cis- conformer with both
charge models at 300 K. Given the different spectroscopic features for
cis- and trans-CysNO, for MbSNO also both conformers were considered
in the MD simulations. The dynamics and spectroscopy of the WT and
S-nitrosylated variants were studied at 50 K and 300 K using PC and
MTP charge models.\\

\begin{figure}[H]
  \centering
    \includegraphics[width=0.75\linewidth]{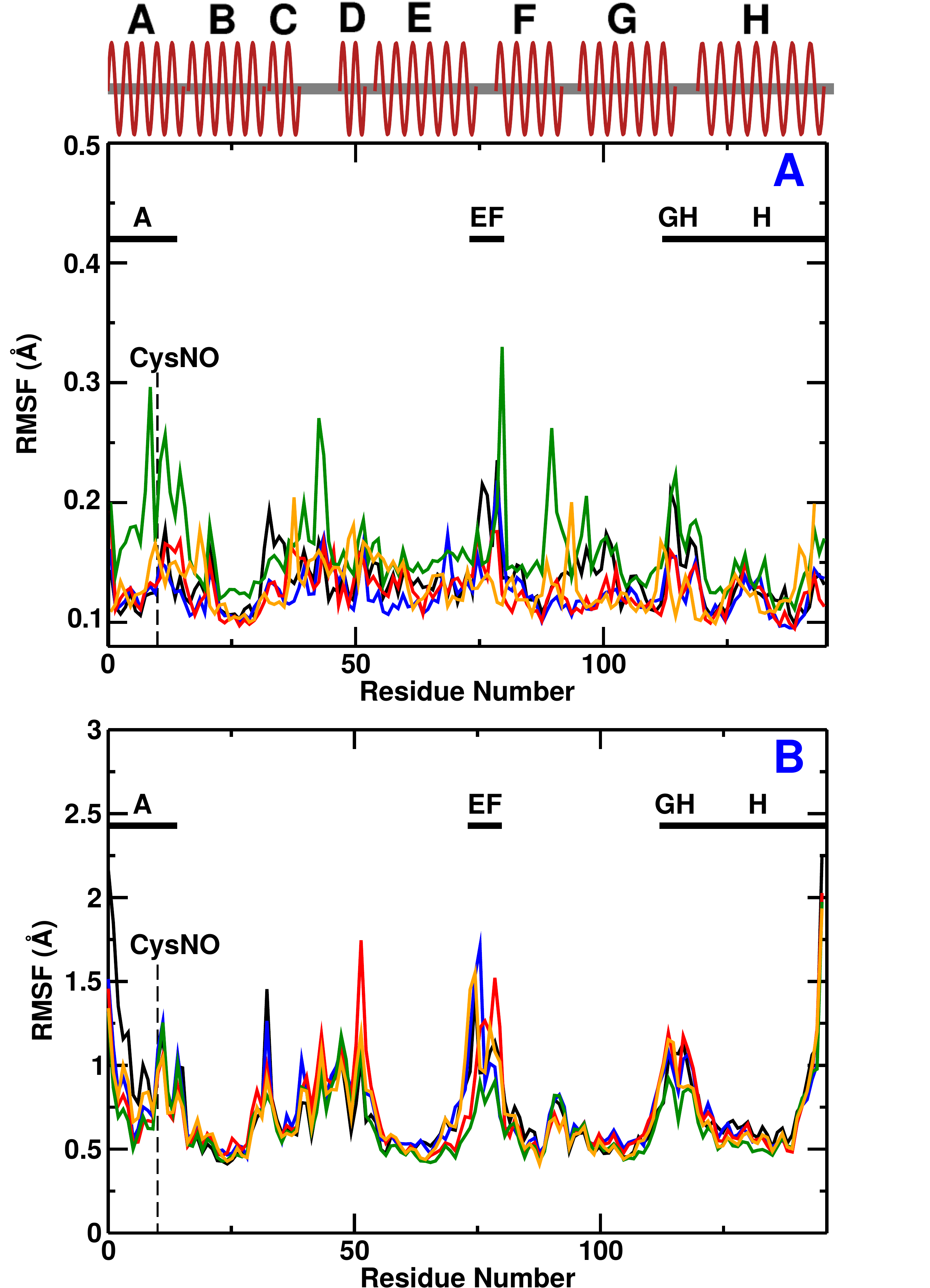}\\
    \caption{Root Mean Squared Fluctuation of each residue at 50 K
      (panel A) and 300 K (panel B) from simulations using the PC and
      MTP model. (Black line: WT, Red line: cis-MbSNO with PC, Blue
      line: trans-MbSNO with PC, Green line: cis-MbSNO with MTP and
      Orange line: trans-MbSNO with MTP). The location of CysNO (at
      position 10), and helix A (Ala2 to Glu15), helix H (Gly121 to
      Ser146), loop EF (Ala74 to Ile81) and loop GH (Glu113 to Gly120)
      are explicitly indicated. Note the different scales along the
      $y-$axis in panels A and B.}
\label{fig:rmsf}
\end{figure}

\noindent
The root mean squared fluctuations of the C$_{\alpha}$ atoms of every
residue at 50 K and 300 K from 10 ns simulations with the PC and MTP
models are reported in Figure~\ref{fig:rmsf} (top and bottom) for the
WT (black), cis-MbSNO (PC, red), trans-MbSNO (PC, blue), cis-MbSNO
(MTP, green), and trans-MbSNO (MTP, orange). At 50 K all RMSFs are
small and S-nitrosylation of Cys10 decreases the flexibility of cis-
and trans-MbSNO with PC in the C-, E-, F-, and G-helix regions
compared with WT. The RMSF for residues Lys31 to Glu35 (end of helix B
and beginning of helix C) display reduced flexibility as a consequence
of the chemical modification. The largest differences are found for
the residues between Lys90 and Leu102 (end of helix F, FG loop,
beginning of helix G). With MTPs on the -SNO label the RMSFs for
cis-MbSNO are typically larger or equal compared with WT (PC) due to
unfavorable conformation around the -SNO group whereas for trans-MbSNO
they are comparable. Contrary to simulations at 50 K that sample the
C$_\alpha $C$_\beta$SN dihedral only at one particular angle, the
dynamics of cis-MbSNO samples states characterized by angles of
$94^{\circ}$ and $80^{\circ}$ (see blue line in
Figure~\ref{sifig:dihedral_50}). The conformational change of
S-nitrosylated Cys10 from $94^{\circ}$ to $80^{\circ}$ at such low
temperature indicates an unfavorable conformation that could lead to a
higher flexibility of the protein.\\

\noindent
At 300 K, the differences in flexibility of each residue between WT,
cis- and trans-MbSNO with both PC and MTP are much smaller. While the
magnitude of the RMSF between 50 K and 300 K increases considerably,
as expected, the differences between the systems become more
specific. The flexibility of residues Ala2 to Cys10 (helix A)
decreased with the modification in both cis- and trans-MbSNO compared
to WT which indicates a rigidification of helix A and points to a
clear impact of nitrosylation at Cys10 on the local dynamics of the
protein. Moreover, increased flexibilities up to 1.7 \AA\/ are found
for residues Glu70 to Ile81 (EF loop). This is the region with the
largest displacements in both, cis- and trans-MbSNO compared to WT.\\

\begin{figure}[H]
  \centering \includegraphics[scale=0.70]{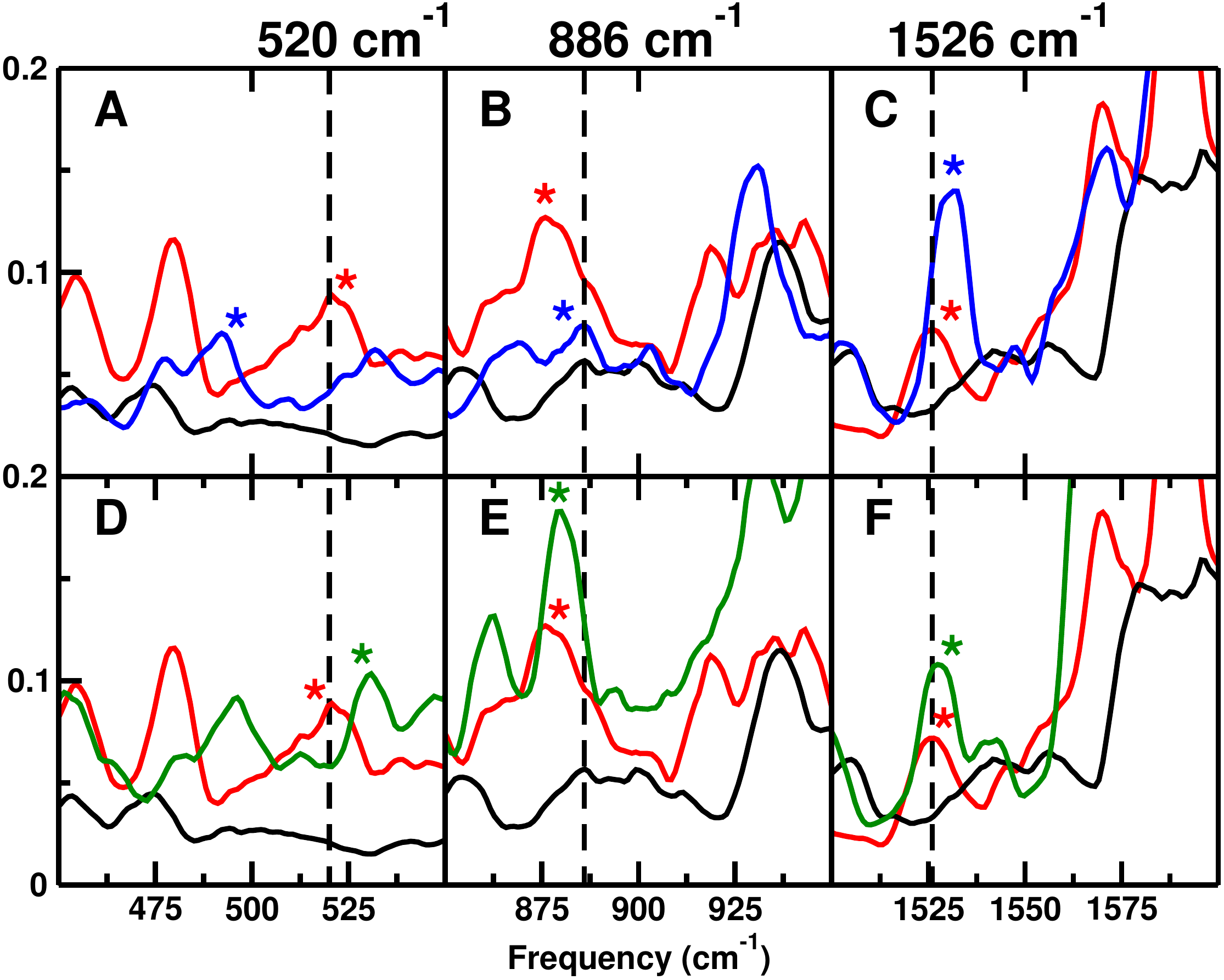}
    \caption{Comparison of the calculated IR spectra for WT, cis- and
      trans-MbSNO from 10 ns simulations at 50 K in the region of the
      SN stretch (panels A and D), SNO bend (panels B and E), and NO
      stretch (panels C and F). Top row from simulations with PC
      (black : WT, red : cis-MbSNO, blue : trans-MbSNO). Bottom row
      from simulations with PC and MTP (black : WT, red : cis-MbSNO
      with PC, green : cis-MbSNO with MTP). The dashed vertical lines
      indicate the experimental values of the corresponding vibrations
      in GSNO\cite{walsh2007s} and the colored stars label the
      spectral signatures to which the mode was assigned based on
      analysis of the power spectra. In general, the WT spectra
      (black) are featureless in the region where the -SNO label shows
      spectroscopic signatures.}
    \label{fig:mb.ir}
\end{figure}

\noindent
The spectra related to the NO and SN stretches and the SNO bend are
reported in Figures~\ref{fig:mb.ir}A to C from simulations for cis-
and trans-SNO with PCs. A comparison for PC and MTP simulations is
given in Figures~\ref{fig:mb.ir}D to F. For the NO stretch
(Figure~\ref{fig:mb.ir}C) the peak for cis-MbSNO from simulations with
PCs appears at 1526 cm$^{-1}$ compared with 1532 cm$^{-1}$ for
trans-MbSNO. These absorption frequencies for the NO stretch compare
with 1526 cm$^{-1}$ from experiments on GSNO in its solid state,
i.e. identical to that for cis-MbSNO, and a blue shift of 6 cm$^{-1}$
for trans-MbSNO between simulations and experiment. It is also noted
that the NO-stretch vibration in CysNO (at 1538 cm$^{-1}$ and 1540
cm$^{-1}$ for cis- and trans-CysNO with MTPs) differs from that for
cis-MbSNO and trans-MbSNO by 14 and 6 cm$^{-1}$, respectively, see
Figure \ref{fig:IR_cys}. Likewise, a blue-shift was observed for
trans-CH$_{3}$SNO relative to cis-CH$_{3}$SNO for experiments in
argon.\cite{mueller1984two} The intensity of the NO band for
trans-MbSNO is higher compared with that of
cis-MbSNO. Figure~\ref{fig:mb.ir}C also shows that the NO stretch is
clearly set apart from the nearby the amide-II band which is ranging
from 1500 to 1620 cm$^{-1}$ in the simulations and from 1500 to 1600
from experiments.\cite{van2001fourier} This should make it possible to
locate the NO stretch mode from experiments, in particular when
reference spectra for the WT protein or for two different isotopes of
the label ($^{14}$NO and $^{15}$NO) are subtracted.\\

\noindent
The SN stretch and SNO bend modes are more challenging to identify for
both conformations. The spectrum in Figure~\ref{fig:mb.ir}B has the
SNO bending mode at 876 cm$^{-1}$ for cis-MbSNO which shifts to 886
cm$^{-1}$ for the trans-conformer. Finally, the SN stretch appears at
522 cm$^{-1}$ for cis-MbSNO and at 494 cm$^{-1}$ for the
trans-conformer, see Figure~\ref{fig:mb.ir}A. Hence, the orientation
of the NO group (cis vs. trans) shifts these modes by 6 to 28
cm$^{-1}$. Again, the magnitude of these shifts is consistent with the
findings for CysNO in water.\\

\noindent
In order to quantify the influence of the charge model used on the
infrared signatures of the SN stretch, SNO bending, and the NO
stretch, cis-MbSNO was considered. For this, a 10 ns MD simulation at
50 K was carried out with MTPs on the -SNO moiety. The corresponding
spectra are shown in Figures~\ref{fig:mb.ir}D to F. The NO stretch
from the simulations with MTP appears at 1527 cm$^{-1}$ which is
shifted by 1 cm$^{-1}$ to the blue compared with the simulation with
the PC model. The intensity of this peak increases considerably when
using the more elaborate model for the electrostatics, see
Figure~\ref{fig:mb.ir}F. For the SNO bending vibration the frequency
maximum appears at 880 cm$^{-1}$ in the simulation with MTPs which is
a shift of 4 cm$^{-1}$ to the blue compared with the PC
simulation. Finally, the SN stretch is at 531 cm$^{-1}$, shifted by 9
cm$^{-1}$ to the blue compared with the PC simulation. Hence, all
three vibrations shift to the blue in simulations with the MTP model
compared with PCs. Such shifts are typical for simulations with PCs
and
MTPs.\cite{MM.mbco:2003,MM.stark:2009,MM.mtp:2009,MM.ices:2008,MM.mbco:2008}
The increased intensity with MTP model made it easier the detect the
vibrations on the total IR spectrum of the protein, especially for SNO
and NO. \\

\noindent
In summary, it is noted that in the regions of the SN stretch, SNO
bend and NO stretch modes the intensity of the spectrum for WT
myoglobin (black trace in Figure \ref{sifig:mb.ir}) is
characteristically low whereas those for the two nitrosylated variants
show increased intensities (see arrows in Figure
\ref{sifig:mb.ir}). The calculated IR spectra from simulations with
PCs and the Fourier transform of the protein dipole autocorrelation
function for WT, cis-MbSNO with PC and trans-MbSNO are reported in
Figure \ref{sifig:mb.ir}. Spectral features of amide-I, amide-II and
amide-III absorption bands are present in all spectra with varying
intensities. Also, the blue shifted frequencies in particular for the
NO-stretch vibration for trans- versus cis-SNO is consistent with
experiments on model compounds in the gas phase or in argon
matrices.\\

\subsection{Water Structure and Global Structural Changes}
\begin{figure}[H]
  \centering
  \includegraphics[width=0.7\linewidth]{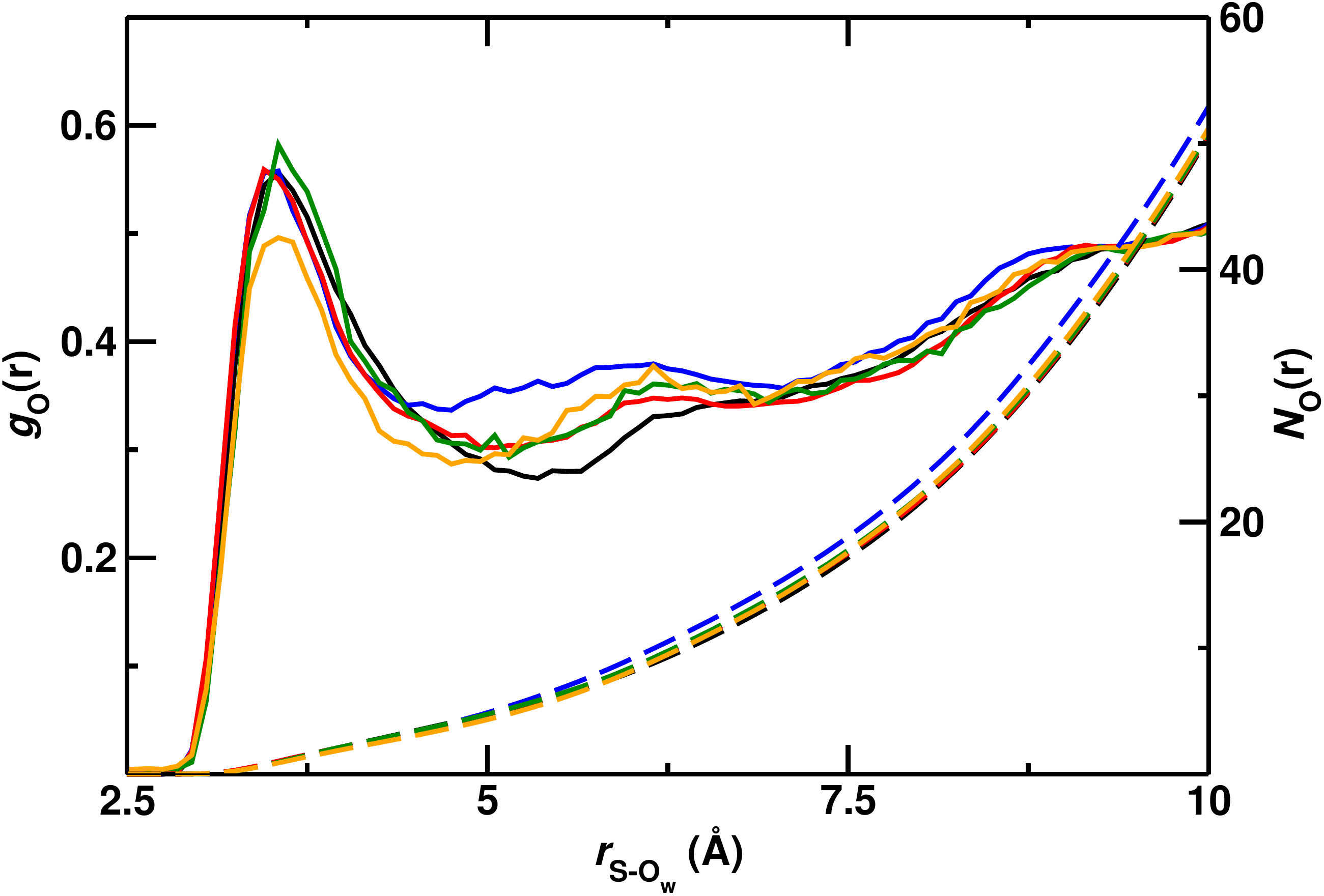}
    \caption{Radial distribution function of water oxygen and the
      corresponding coordination number of $N_{\rm O}(r)$ of water
      oxygen with respect to the sulfur atom obtained from $NVE$
      simulations at 300 K. Color code: WT (black), cis-MbSNO with PC
      (red), trans-MbSNO with PC (blue), cis-MbSNO with MTP (green),
      trans-MbSNO with MTP (orange).}
    \label{fig:radial}
\end{figure}

Next, it is of interest to consider the local water ordering around
the modification site (Cys10) for WT and nitrosylated Mb. The radial
distribution function $g_{\rm S-OW}(r)$ and corresponding coordination
number $N_{\rm S-OW}(r)$ of water oxygen (OW) with respect to the
sulfur atom of Cys10 in WT, cis-MbSNO, and trans-MbSNO are shown in
Figure~\ref{fig:radial}. At 300 K and with PC for the simulations the
first solvation shell peak appears at 3.5 \AA\ in all proteins. While
the difference of the first solvation peak in terms of peak height and
shape is marginal, the radial distribution functions for cis- and
trans-MbSNO (red and blue) differ from that for WT (black) in the
region between 4 \AA\/ and 7.5 \AA\/, see Figure~\ref{fig:radial}. It
clearly shows that with respect to the Cys10-sulfur atom both,
cis-MbSNO and trans-MbSNO, are more highly solvated in the range of
$\sim 5$ to 7 \AA\/ compared to WT. Since nitric oxide is more solvent
exposed in trans-MbSNO (blue trace) compared to cis-MbSNO (red trace)
the -SNO label is more highly hydrated in the trans-conformation. In
terms of the overall hydration, the coordination number of WT and
trans-MbSNO differs by 3 water molecules at a distance of 10 \AA\/,
see Figure~\ref{fig:radial}, whereas WT and cis-MbSNO have the same
overall hydration. Although the -SNO (modified Mb) group requires more
space compared with -SH (WT Mb), attaching the -NO label recruits more
water molecules. This is evidently an effect of the different
electrostatics between water and the -SH and -SNO groups,
respectively.\\

\noindent
With the MTP model, trans-MbSNO displays a shallower first solvation
shell compared to PC, while cis-MbSNO with MTP has comparable
solvation to PC. The dissimilar impact of additional multipoles on the
two conformers can be rationalized by the
$\phi(\rm{NC_{\alpha}C_{\beta}S})$ angles sampled during the
dynamics. For cis-MbSNO and both charge models this angle fluctuates
around $\sim -60^{\circ}$ throughout the 10 ns simulation (see red and
green lines in Figure~\ref{sifig:dihedral_300}B). However, for
trans-MbSNO (see Figure~\ref{sifig:dihedral_300}B) the
$\phi(\rm{NC_{\alpha}C_{\beta}S})$ angle sampled differs between PC
and MTP. With both models $\phi(\rm{NC_{\alpha}C_{\beta}S})$ samples
structures with $\sim -55^{\circ}$ - albeit only briefly for PCs - and
$\sim 52^{\circ}$. For the simulation with PCs, in addition the
orientation with $\phi(\rm{NC_{\alpha}C_{\beta}S}) = -170^{\circ}$ is
sampled. Also, the residence time in the configuration with
$\phi(\rm{NC_{\alpha}C_{\beta}S}) \sim -55^{\circ}$ differs between
simulations with PC (9 ns) compared with MTP (5.75 ns) and the
simulation with PCs spontaneously returns to the conformation with
$\phi(\rm{NC_{\alpha}C_{\beta}S}) = -62^{\circ}$. Consequently, the
different level of solvent exposure can be the underlying reason for
the shallower solvation shells observed with MTP. However, longer
simulations (and hence more transitions) are required for quantitative
characterizations of such residence times.\\

\begin{figure}[H]
  \centering
  \includegraphics[width=0.47\linewidth]{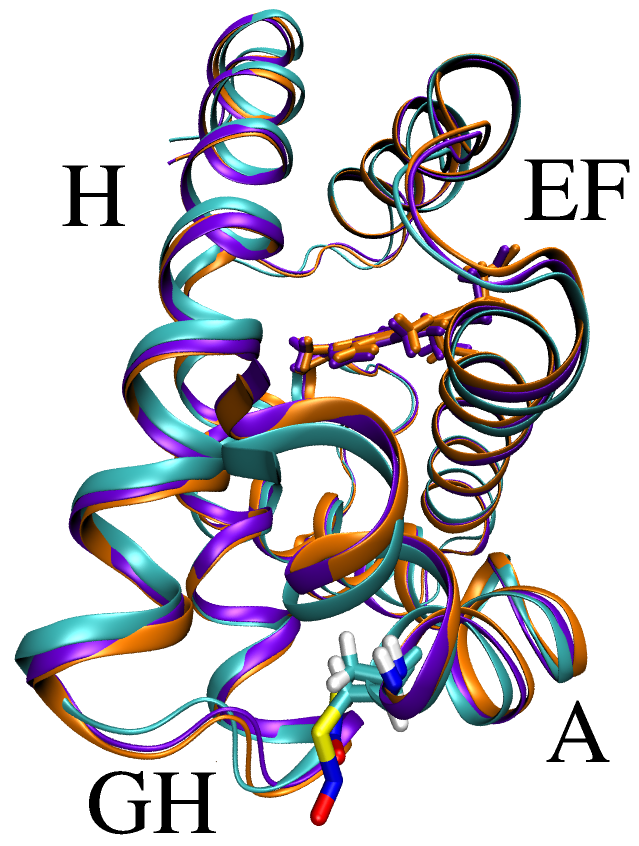}
  \includegraphics[width=0.49\linewidth]{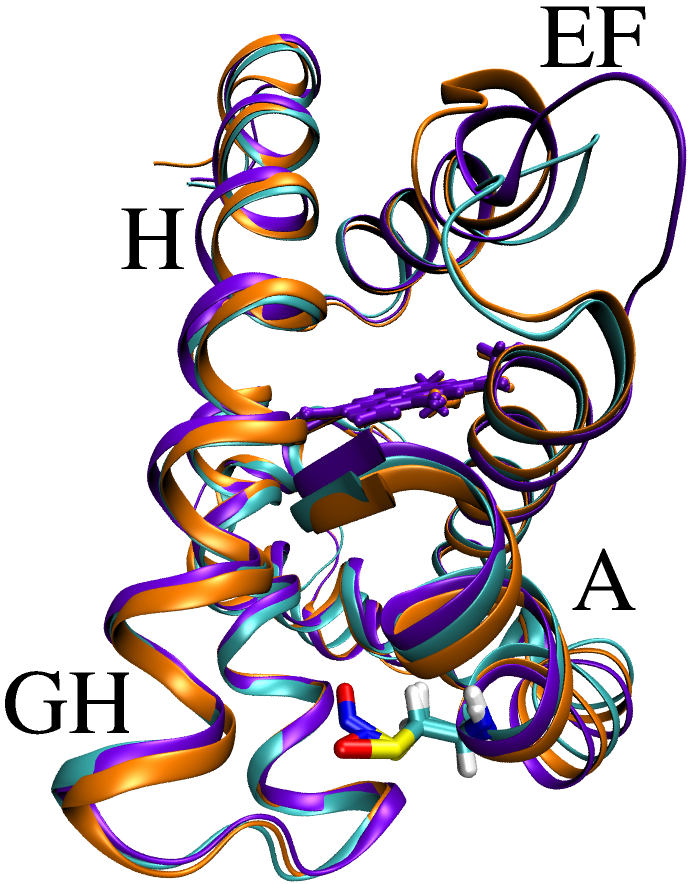}
    \caption{The conformational changes in the protein structure
      induced by S-Nitrosylation, averaged over the last ns of free
      dynamic simulations at 50 (Left) and 300 K (Right) with PC
      model. WT (cyan), cis-MbSNO (orange) and trans-MbSNO
      (violet). S-Nitrosylated Cys10 and the heme unit are represented
      by licorice. A, H helices and EF, GH loops are labeled.}
   \label{fig:conf_change}
\end{figure}

\noindent
{\it Global Changes:} For characterizing more global structural
changes, the last nanosecond of the 10 ns production run have been
analyzed to investigate the structural changes in the protein at 50 K
and 300 K, respectively. The cis-MbSNO and trans-MbSNO structures are
superimposed onto the WT simulation and the structures averaged over
the last ns of the free dynamic simulations at 50 K (left) and 300 K
(right) in Figure~\ref{fig:conf_change} for WT (cyan), cis-MbSNO
(orange), and trans-MbSNO (violet), respectively. The dihedral angles
$\phi(\rm{C_{\beta}SNO})$ (for cis- versus trans-),
$\phi(\rm{C_{\alpha}C_{\beta}SN})$ and $\phi(\rm
N{C_{\alpha}C_{\beta}S})$ (for the position of the NO with respect to
nearby loops and $\alpha-$helices) determine the orientation of the
-SNO label and reported in Figure~\ref{sifig:dihedral_50} for 50 K and
Figure~\ref{sifig:dihedral_300} at 300 K. The time series show that
all states sampled are also observed in the 2NRM X-ray structure.\\

\begin{table}[h]
\centering
\caption{The average C$_{\alpha}$ RMSD (in \AA\/) for Helix A, Helix
  H, Loop GH, Loop EF and the entire protein for cis-MbSNO and
  trans-MbSNO with respect to the WT X-Ray structure for the last
  nanosecond of a 10 ns free dynamics simulation at 50 and 300 K with
  PC and MTP models. As a comparison, the differences between the 2NRL
  and 2NRM structure are also given.}
\begin{tabular}{l | c c | c c | c c | c c | c}
\hline
 & \multicolumn{4}{c}{50K} & \multicolumn{4}{c}{300K} & \multicolumn{1}{c}{X-Ray}\\
\hline
 &  \multicolumn{2}{c}{cis-MbSNO}& \multicolumn{2}{c}{trans-MbSNO}& \multicolumn{2}{c}{cis-MbSNO}& \multicolumn{2}{c}{trans-MbSNO} & 2NRM\\
\hline
& PC & MTP & PC  & MTP & PC & MTP & PC & MTP & \\
\hline
Helix A & 0.57 &0.60 &0.51 &0.63 & 1.45& 1.48 & 1.37 &1.33 & 0.681 \\
Helix H & 0.86 &1.67& 1.00 &1.01 & 1.19 &1.19 & 1.11 &0.86  & 0.627\\
Loop GH & 1.31 &0.71 & 0.68 &0.93 & 1.72 &1.68& 1.55 &1.92 & 1.076\\
loop EF & 1.17 &1.07 & 1.16 &1.35 & 2.63 &2.02 & 3.23 &2.54  & 3.843\\
Protein & 0.83 &0.78& 0.81 &0.80  & 1.27 &1.14 & 1.28 &1.17  & 0.989 \\
\hline
\label{tab:RMSD_table}
\end{tabular}
\end{table}

\noindent
Although the entire protein structure is affected by the modification
to a certain degree, four regions revealed most prominent
changes. They are helix A (residues Ala2 to Glu15), helix H (residues
Gly121 to Ser146), loop GH (residues Glu113 to Gly120), and loop EF
(residues Ala74 to Ile81). The C$_{\alpha}$ RMSD of helix A, helix H,
loop GH, loop EF and the entire protein for cis- and trans-MbSNO with
respect to the WT X-ray structure during the last nanosecond of the 10
ns free dynamics simulations are summarized in
Table~\ref{tab:RMSD_table}.\\

\noindent
At 50 K, the C$_{\alpha}$ RMSD for [cis-MbSNO, trans-MbSNO] compared
to X-Ray WT are [0.83, 0.81] \AA\/ with PC and [0.78, 0.80] \AA\/ with
MTP, respectively, which signals good preservation of the overall
structure for the methods chosen. The C$_{\alpha}$ RMSD calculated
between the 2NRL and 2NRM structures is 0.99 \AA\/ which is compatible
with the C$_{\alpha}$ RMSD calculated from the simulations with both,
PC and MTP electrostatic models for the NO-label. These results show
that the orientation of nitric oxide has a limited effect on the
overall structure of the protein with only 0.02 \AA\ difference
between the two conformers with both charge models. Further, the same
trend is observed in the local dynamics of the helices. C$_{\alpha}$
RMSD values of the individual helices (A and H) deviate by 0.06
\AA\ to 0.14 \AA\ for cis-MbSNO compared to trans-MbSNO with PC,
respectively. These small but consistent differences in the RMSD
values between the two conformers can be rationalized by the
orientation of nitric oxide with respect to $\phi(\rm{C_{\beta}SNO})$
which is 0$^{\circ}$ in cis-MbSNO and 180$^{\circ}$ in trans-MbSNO. \\

\noindent
At 300 K, the C$_{\alpha}$ RMSD of the cis-MbSNO and trans-MbSNO with
respect to the WT X-Ray structure are 1.27 \AA\/, 1.28 \AA\/ with PC
and 1.14 \AA\/, 1.17 \AA\/ with MTP, respectively.  The local motion
of the protein causes larger C$_{\alpha}$ RMSD values of helix A in
both, trans-MbSNO with PC and cis-MbSNO with PC. The values, with
respect to X-Ray WT structure, were 1.45 \AA\ and 1.37 \AA\,
respectively. The C$_{\alpha}$ RMSD of helix H and loop GH were 1.19
and 1.72 \AA\ in cis-MbSNO with PC, compared with 1.11 and 1.55
\AA\ in trans-MbSNO with PC. Furthermore, the largest deviation from
the X-Ray WT structure is observed for the EF loop, see Table
\ref{tab:RMSD_table}. The relatively flexible loop moved closer to
helix H in cis-MbSNO. Contrary to that, the movement of the loop was
in the opposite direction for trans-MbSNO with PC. The C$_{\alpha}$
RMSD of loop EF is 2.63 \AA\ in cis-MbSNO with PC and 3.23 \AA\/ in
trans-MbSNO with PC.\\

\begin{figure}[H]
  \centering \includegraphics[width=0.40\linewidth]{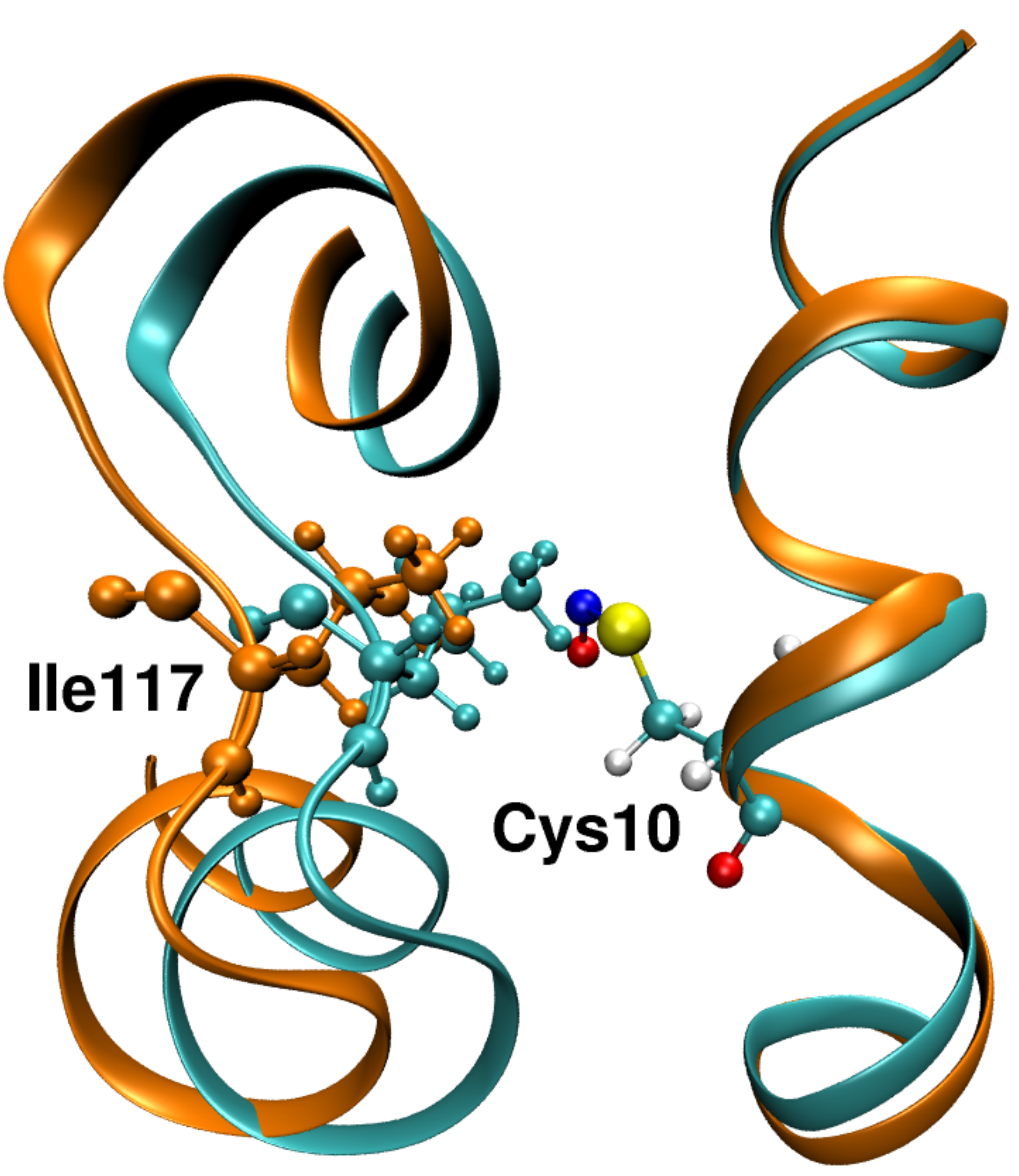}
    \caption{CPK representation of S-nitrosylated Cys10 and Leu117
      (labelled) which demonstrates the crowding that would occur if
      there was no helix movement in cis-MbSNO. The WT and cis-MbSNO
      structures are shown by cyan and orange cartoon representations,
      respectively.}
    \label{fig:crowding}
\end{figure}

\noindent
Depending on the orientation of the -SNO group (cis or trans),
additional contacts with the protein can emerge. At 50 K, the
$\phi(\rm{C_{\alpha}C_{\beta}SN})$ angles sampled for cis- and
trans-MbSNO are shown in Figure~\ref{sifig:dihedral_50}. For cis-MbSNO
the -SNO group is closer to the GH loop than for
trans-MbSNO. Consequently, nitrosylation of Cys10 in its cis-conformer
will lead to steric hindrance with the GH loop and pushes it away from
its position in WT Mb to avoid overlap between Leu117 and
Cys10. Concomitantly, helix A (containing residue Cys10) is forced in
the opposite direction of loop GH. Finally, helix H can push into the
void created by the movement of helix A. This same steric hindrance
was present to a lesser extent for trans-MbSNO and WT. The movement of
the EF loop is also visible in Figure~\ref{fig:conf_change}. The loop
had a C$_{\alpha}$ RMSD of 1.17 \AA\/ in cis-MbSNO and 1.16 \AA\/ in
trans-MbSNO. The resulting crowding involving residues Cys10 and
Leu117 without concomitant motion of loop GH and helix A upon
nitrosylation is shown in Figure~\ref{fig:crowding}.\\

\noindent
At 300 K, the influence of the NO-modification on the structure of Mb
is more pronounced than for 50 K, specifically for loop EF. However,
the similar C$_{\alpha}$ RMSD for cis- and trans-MbSNO with respect to
WT (see Table ~\ref{tab:RMSD_table}) shows that the orientation of the
nitric oxide has only a limited effect on the overall structure of the
protein. In the conformation with $\phi(\rm{C_{\alpha}C_{\beta}SN})
\sim -85^{\circ}$ (for cis-MbSNO with PC (red) and trans-MbSNO with
MTP (orange), see Figure \ref{sifig:dihedral_300}A) the nitric oxide
group resides midway between loop GH and helix A. The conformational
transition to $\phi(\rm{C_{\alpha}C_{\beta}SN}) \sim 160^{\circ}$
moves the -NO group closer to helix H. For cis-MbSNO with MTPs (green)
this is the only state sampled throughout the simulation and
corresponds to the major component observed in the 2NRM X-ray
structure. For cis-MbSNO with PCs (red) there is a spontaneous
transition between the minor and the major conformer. This suggests
that the major conformer is probably lower in energy but for a firm
conclusion on this considerably more extended simulations are
required.\\

\noindent
For trans-MbSNO (blue and orange traces in Figure
\ref{sifig:dihedral_300}A) four metastable states for the orientation
of $\phi(\rm{C_{\alpha}C_{\beta}SN})$ were found in simulations with
PCs. The orientation with $\phi(\rm{C_{\alpha}C_{\beta}SN}) \sim
155^{\circ}$ is prevalent (Figure~\ref{sifig:dihedral_300}A). In this
conformation the NO modification faces towards helix A. After 1 ns, a
transition to $\phi(\rm{C_{\alpha}C_{\beta}SN}) \sim 52^{\circ}$
occurs. Concomitantly, the $\phi(\rm N{C_{\alpha}C_{\beta}S})$
dihedral changes from $\sim -55$ to $\sim -180^{\circ}$
(Figure~\ref{sifig:dihedral_300}B) which positions the nitric oxide
towards the GH loop. A next transition leads to
$\phi(\rm{C_{\alpha}C_{\beta}SN}) \sim -70^{\circ}$ with the NO midway
between loop GH and helix A, as was found for cis-MbSNO. Finally, a
transition to a state with $\phi(\rm N{C_{\alpha}C_{\beta}S}) \sim
55^{\circ}$, and $\phi(\rm{C_{\alpha}C_{\beta}SN}) \sim 95^{\circ}$
occurred in which the NO faced towards the inner part of the
protein. One should note that this was the only occasion on which SNO
was buried into the protein. This conformation reduces the probability
for the solvent accessibility of SNO.\\

\noindent
S-nitrosylation-induced structural changes have also been reported for
the crystal structures of the WT (2NRL) and cis-MbSNO (2NRM). This is
exacerbated by the high C$_{\alpha}$ RMSD of loop EF as was observed
in the present simulations (see Table ~\ref{tab:RMSD_table}). X-ray
experiments revealed that the CysNO can induce crowding between Leu117
(loop GH) and Ala6 (helix A) if these parts of the protein structure
did not move\cite{schreiter2007s} which also been observed in the
simulation of cis-MbSNO. The distance between the C$_{\alpha}$ atoms
of Cys10 and Leu117 ($r_{10-117}$) in the X-Ray structure is 5.97
\AA\/ (for 2NRL, WT) and 6.35 \AA\/ (for the major conformer of
2NRM). This compares with an averaged $r_{10-117}$ of 6.18 \AA\ for WT
(black trace in Figure~\ref{sifig:ca_dist_50}) and 7.39 for cis-MbSNO
with PC (red trace in Figure~\ref{sifig:ca_dist_50}) in the
simulations. The same simulation for cis-MbSNO with MTP leads to
relaxation of the structure after $\sim 2$ ns (green trace in
Figure~\ref{sifig:ca_dist_50}). Finally, restarting the cis-MbSNO
simulation with PC after 10 ns but with MTPs on the -SNO group also
leads to relaxation towards the value from the X-ray structure. So,
with MTP model, $r_{10-117}$ assumed the value found in X-Ray whereas
with PC it does not. This finding shows that repositioning of loop GH
and helix A occurs to accommodate residues Cys10 and Leu117 in
cis-MbSNO. The effect is demonstrated in
Figure~\ref{fig:crowding}. Although the averaged $r_{10-117}$
decreases to 6.37 \AA\ for cis-MbSNO with MTP, the steric overlap is
prevented due to sampling the $\phi(\rm{C_{\alpha}C_{\beta}SN})$
dihedral at different angles ($85^\circ$ and $94^{\circ}$) than
cis-MbSNO with PC (red in Figure \ref{sifig:dihedral_50}). Quantum
chemical calculations indicate that the potential energy curve along
the CCSN dihedral is flat between $50^\circ$ and $300^\circ$ with
minima at $100^\circ$ and $250^\circ$ and a barrier between them of 2
kcal/mol at the MP2 level of theory.\\

\noindent
The structural changes induced by nitrosylation at Cys10 also
influence local hydration between the helix A, helix H and loop GH,
see Figure \ref{sifig:hydration}. The chemical modification leads to a
decrease of hydration in this region by 30 \% between the WT and
trans-MbSNO. Given the prominent role that water molecules can play in
protein folding,\cite{weik:2015} and for
function,\cite{pocker:2000,zewail:2004} such a change in hydration may
also be functionally relevant for a PTM such as nitrosylation. Also,
the degree of hydration may affect the stability of the protein as has
recently been demonstrated for insulin dimer. Mutation of residue
PheB24 to Ala or Gly leads to water influx and destabilization of the
dimer by a factor of 2 to
3.\cite{Strazza1985,MM.ins:2019,MM.insulin:2018,MM.ins:2005,MM.ins:2004}\\

\section{Conclusion}
The present work reported on the structural, dynamical and
spectroscopic implications of nitrosylation at cysteine. For this,
CysNO as a model and nitrosylated Mb (MbSNO) were considered. For both
systems it was found that cis- and trans-orientations can be
spectroscopically distinguished. While for CysNO the spectroscopic
signatures can be more readily differentiated from other vibrational
modes this is more challenging for MbSNO due to the larger number of
vibrations and overlap with other vibrational
excitations. Nevertheless, the spectroscopic signatures can be clearly
located with both, simulations using PC and MTP models, see Figure
\ref{fig:mb.ir}. From recording and subtracting the IR spectrum for WT
Mb, or the difference spectrum between MbS$^{14}$NO and MbS$^{15}$NO
(as for MbNO\cite{Coyle03}) it should be possible to identify the IR
signatures for nitrosylation in all proteins under physiological
conditions for which such experiments are possible.\\

\noindent
One consequence of nitrosylation is that local hydration changes
around the modification site compared with the WT protein and
different hydration between the A- and H-helices. For the protein
considered here, myoglobin, 3 more water molecules were found to be
recruited for -SNO compared with -SH for modification at
Cys10. Although this number may appear small, given the functional
role of individual, local water molecules that has been established in
other proteins, this observation may still be functionally
relevant. Furthermore, S-nitrosylation leads to discernible
spectroscopic features in the infrared spectrum of MbSNO and to
structural changes near the modification site.\\

\noindent
It is also of interest to note that for nitrosylated Cysteine cis- and
trans-orientations have been observed experimentally in human
thioredoxin at position Cys69.\cite{weichsel:2007} Also, the
nitrosylated Cys62 was completely buried and points towards the
protein interior which was also found for a short time during the
present simulations for the trans-conformer.\\

\noindent
In summary, S-nitrosylation in myoglobin was found to lead to
detectable spectroscopic features in the infrared. Local hydration is
enhanced as a consequence of the modification and the structure and
dynamics at distant sites in the protein can be changed appreciably
and are consistent with X-ray experiments.\\

\section{Acknowledgment}
This work was supported by the Swiss National Science Foundation
grants 200021-117810, 200020-188724, the NCCR MUST, and the University
of Basel which is gratefully acknowledged.

\bibliography{sno-mb} 

\end{document}


\begin{figure}[H]
  \centering
    \includegraphics[width=0.47\linewidth]{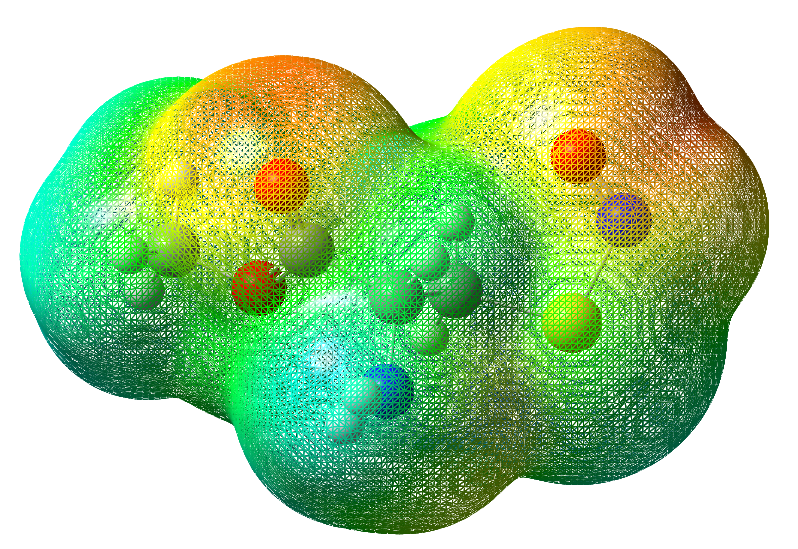}
    \includegraphics[width=0.47\linewidth]{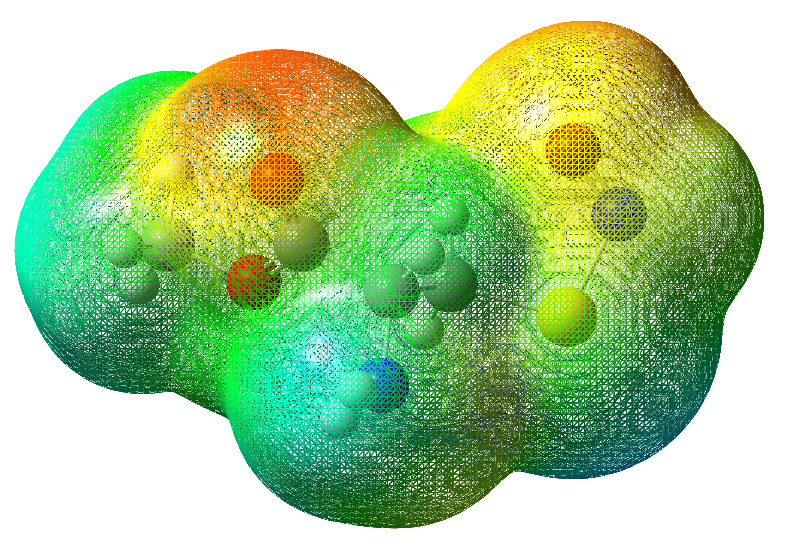}
    \includegraphics[width=0.47\linewidth]{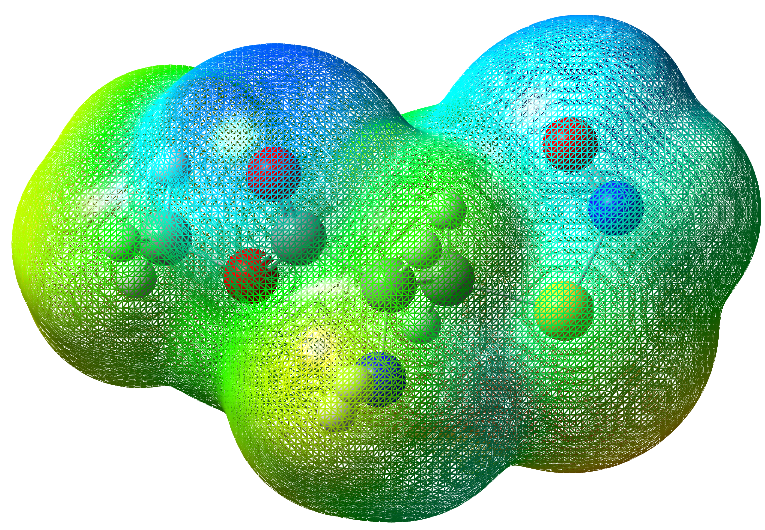}
    \includegraphics[width=0.47\linewidth]{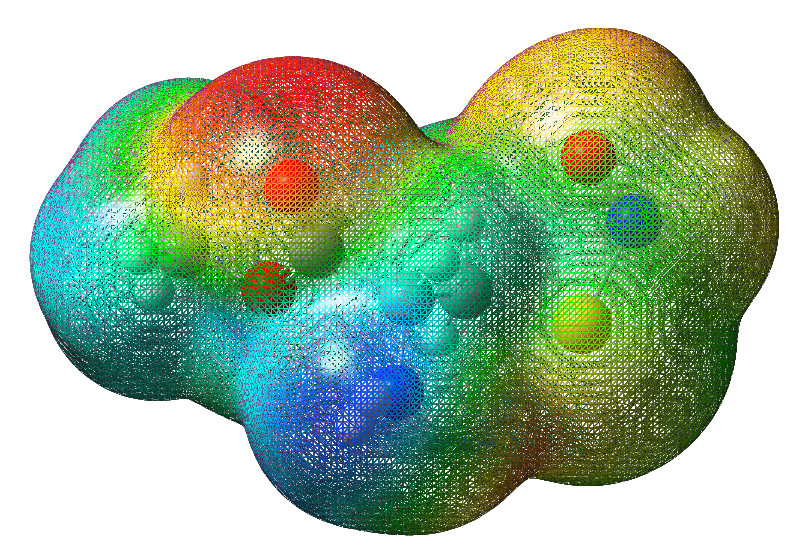}
    \caption{The density cube generated from the cis-CysNO residue
      with PC model (Up-Left) and MTP model (Up-Right). The cube
      generated from the difference between PC and MTP model
      (Down-Left) and reference ESP cube at MP2/aug-cc-pVDZ level
      (Down-Right).}
   \label{sifig:cis_esp}
\end{figure}

\begin{figure}[H]
  \centering
  \includegraphics[width=0.47\linewidth]{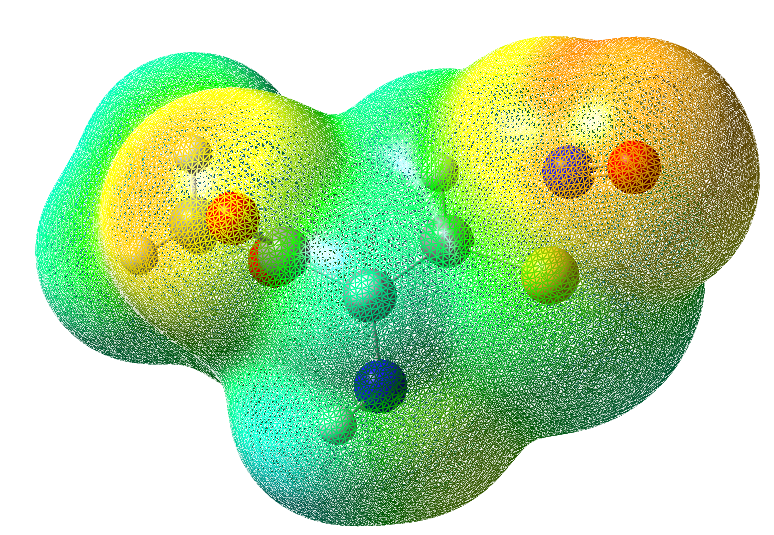}
  \includegraphics[width=0.47\linewidth]{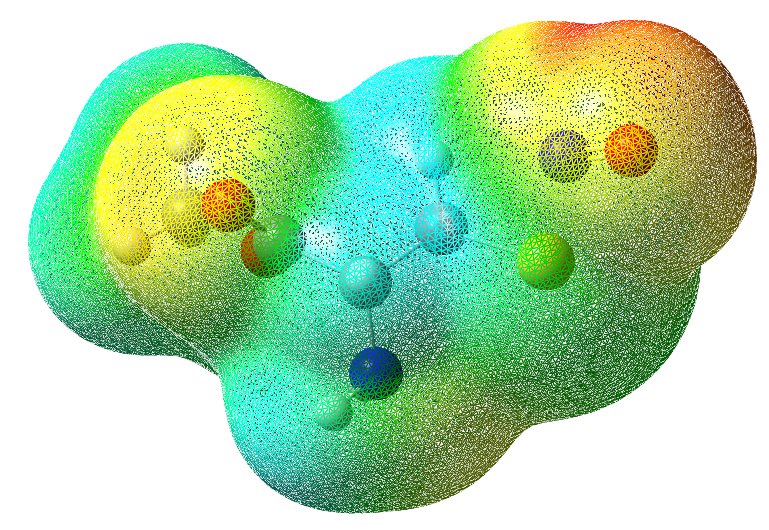}
  \includegraphics[width=0.47\linewidth]{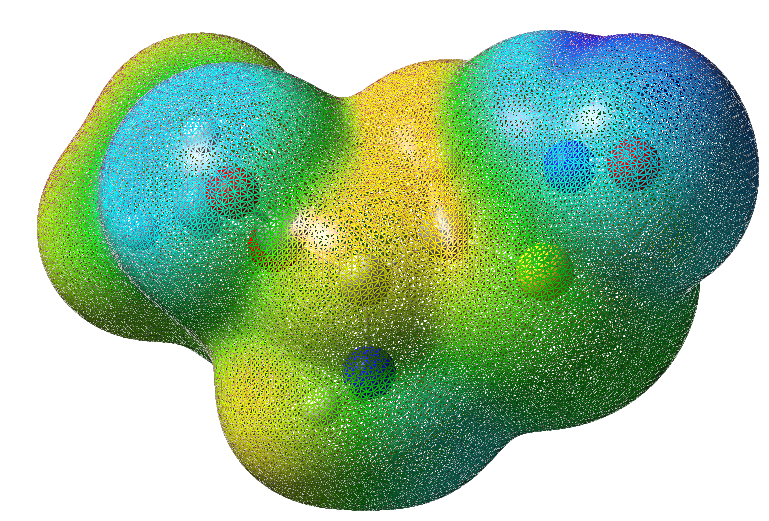}
  \includegraphics[width=0.47\linewidth]{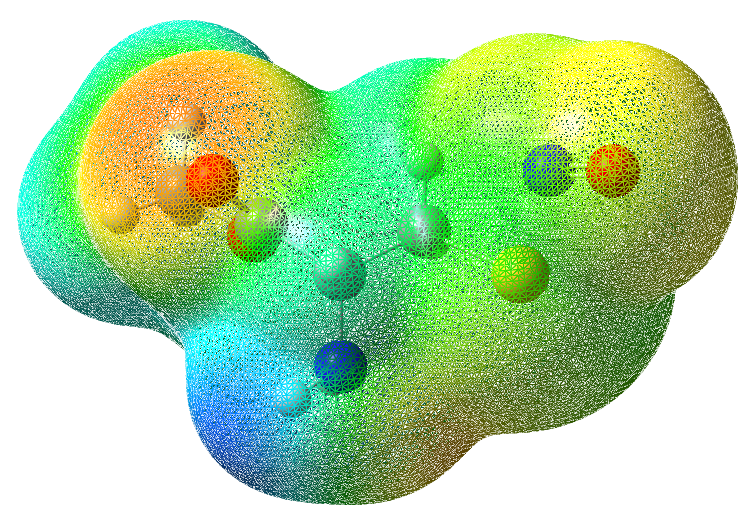}
    \caption{The density cube generated from the trans-CysNO residue
      with PC model (Up-Left) and MTP model (Up-Right). The cube
      generated from the difference between PC and MTP model
      (Down-Left) and reference ESP cube at MP2/aug-cc-pVDZ level
      (Down-Right).}
   \label{sifig:trans_esp}
\end{figure}

\begin{figure}[H]
  \centering
    \includegraphics[width=0.60\linewidth]{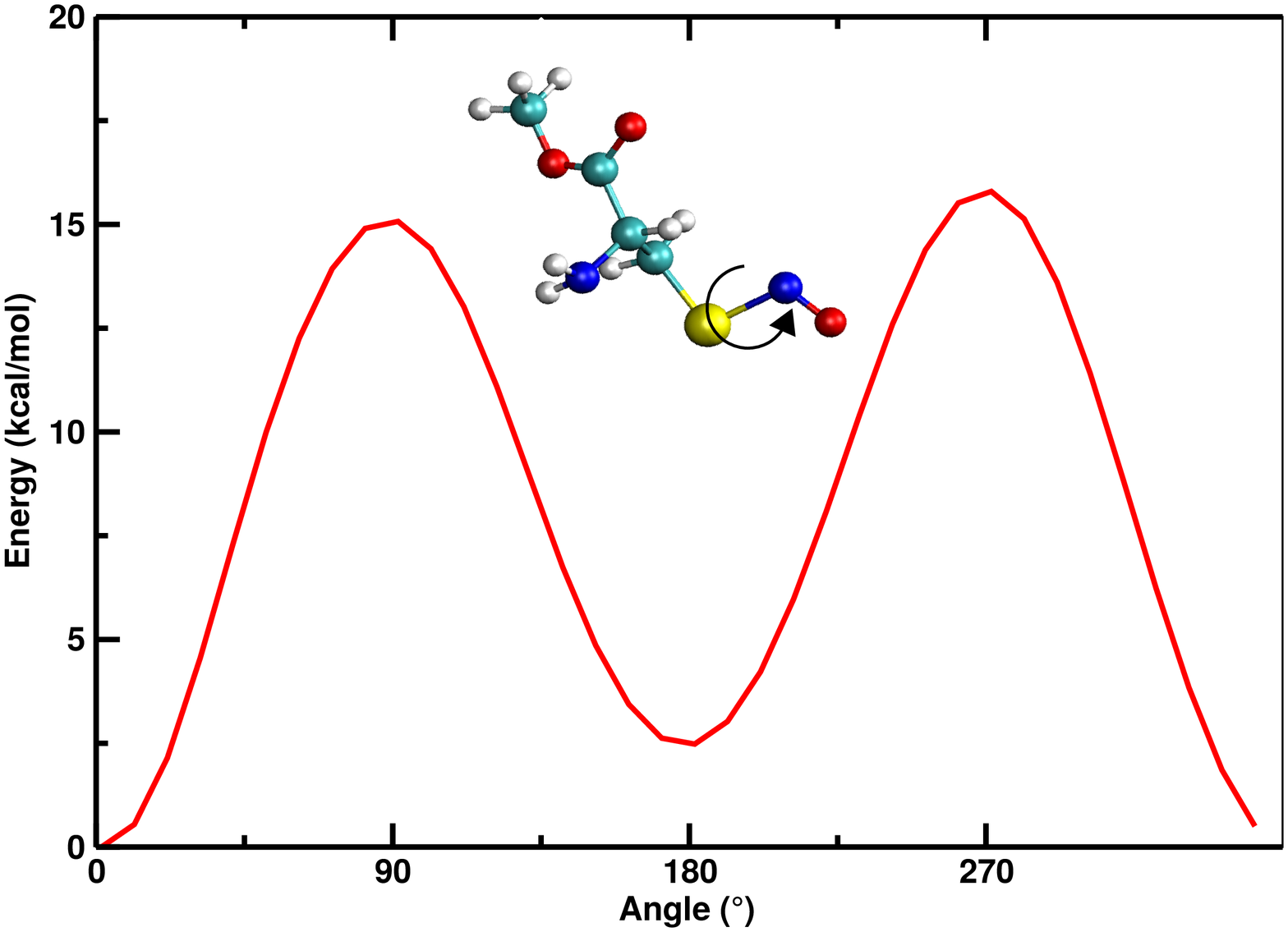}
    \caption{The energy profile of the CSNO torsion angle at the
      MP2/aug-cc-pVDZ level of theory.}
    \label{sifig:csno}
\end{figure}

\begin{figure}[H]
  \centering
  \includegraphics[width=0.49\linewidth]{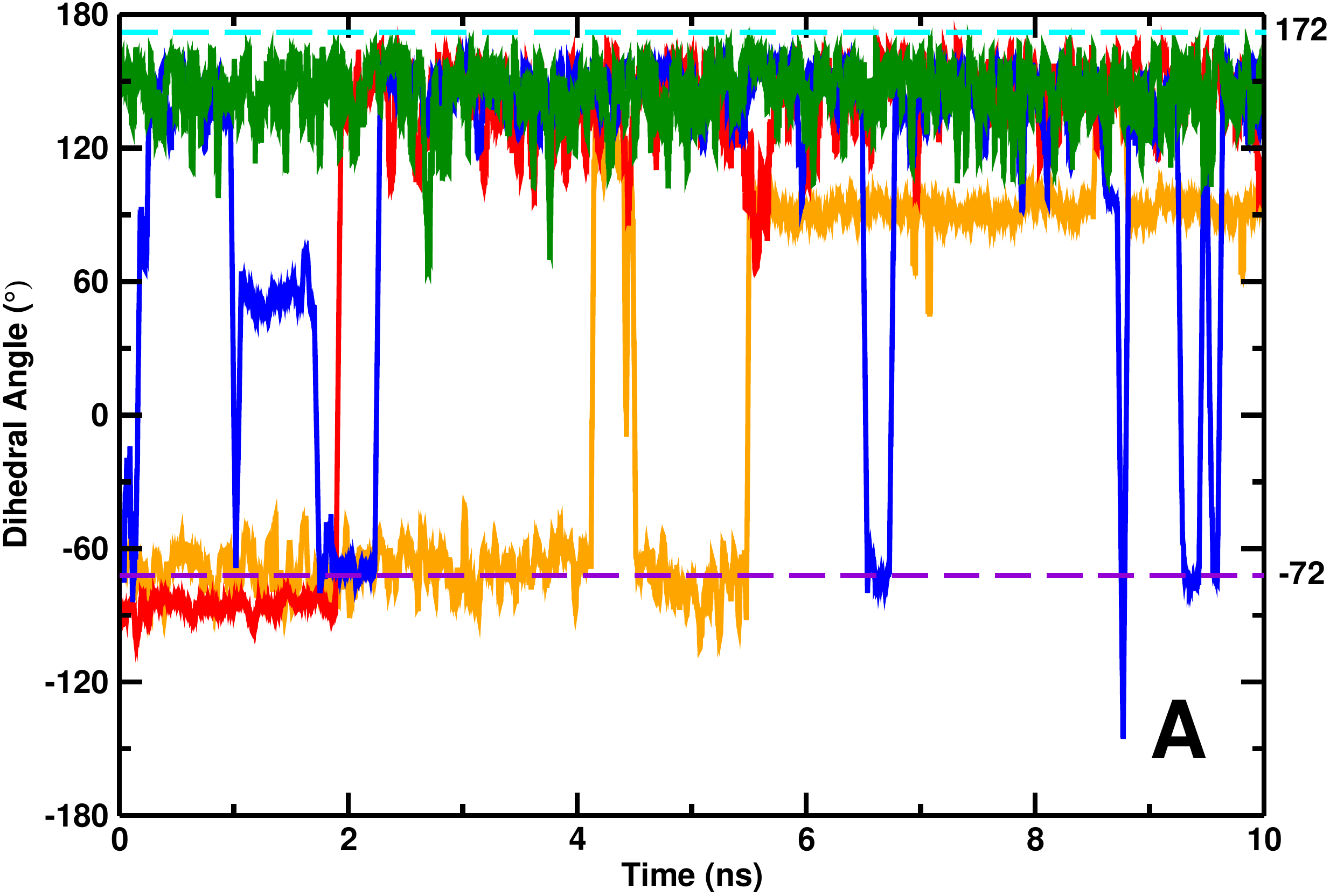}
  \includegraphics[width=0.49\linewidth]{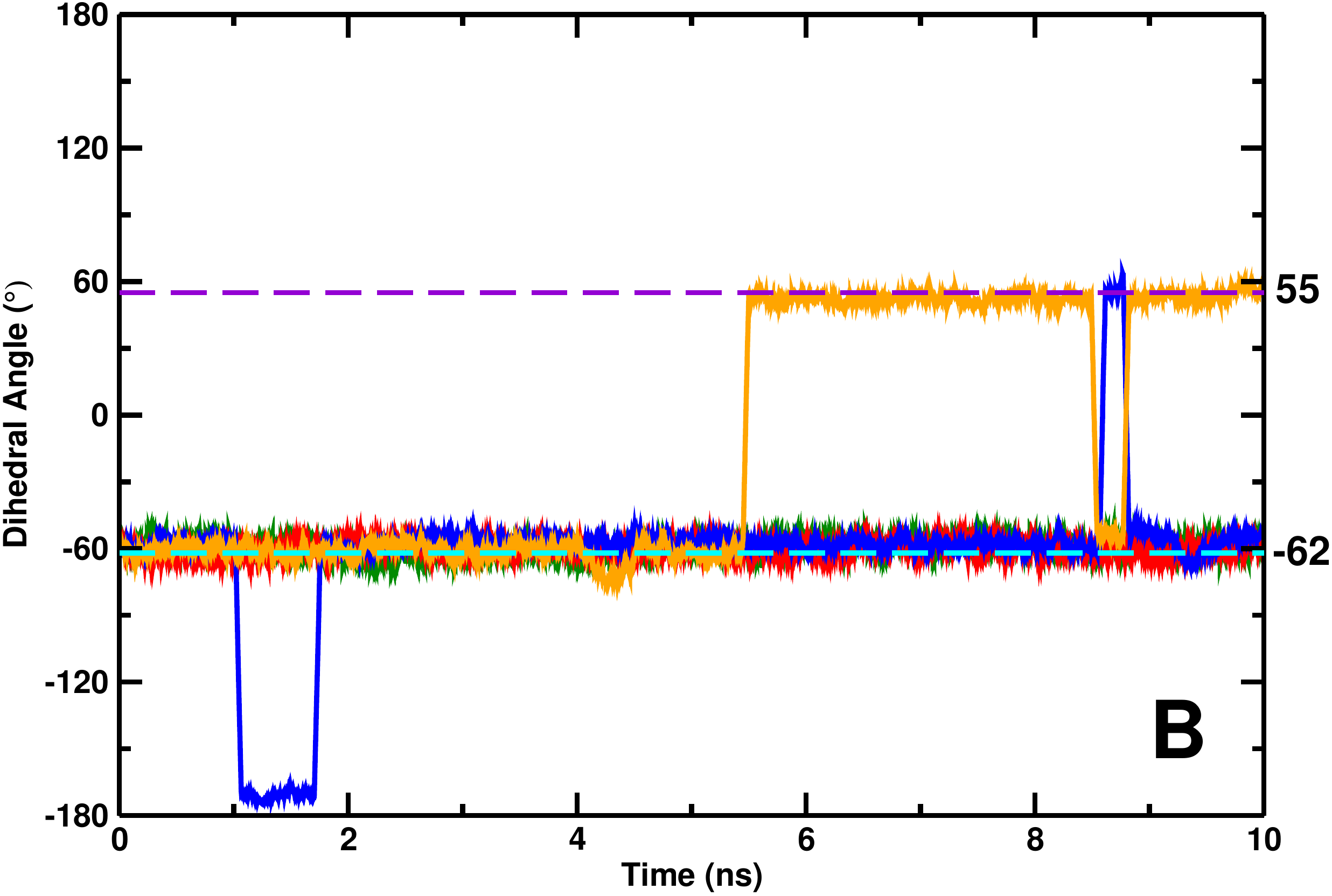}
    \caption{Panel A) Dihedral angle of $\arcangle$C$_\alpha
      $C$_\beta$SN for Cys10 as a function of time in cis-MbSNO with
      PC (red), trans-MbSNO with PC (blue), cis-MbSNO with MTP (green)
      and trans-MbSNO with MTP (orange) at 300 K. Panel B) Dihedral
      angle $\arcangle$NC$_\alpha $C$_\beta$S of Cys10 as a function
      of time in cis-MbSNO with PC (red), trans-MbSNO with PC (blue),
      cis-MbSNO with MTP (green) and trans-MbSNO with MTP (orange) at
      300 K. Dashed lines represent the values from the major (cyan)
      and minor (violet) cis-MbSNO conformer in 2NRM.}
   \label{sifig:dihedral_300}
\end{figure}

\begin{figure}[H]
  \centering
  \includegraphics[width=0.49\linewidth]{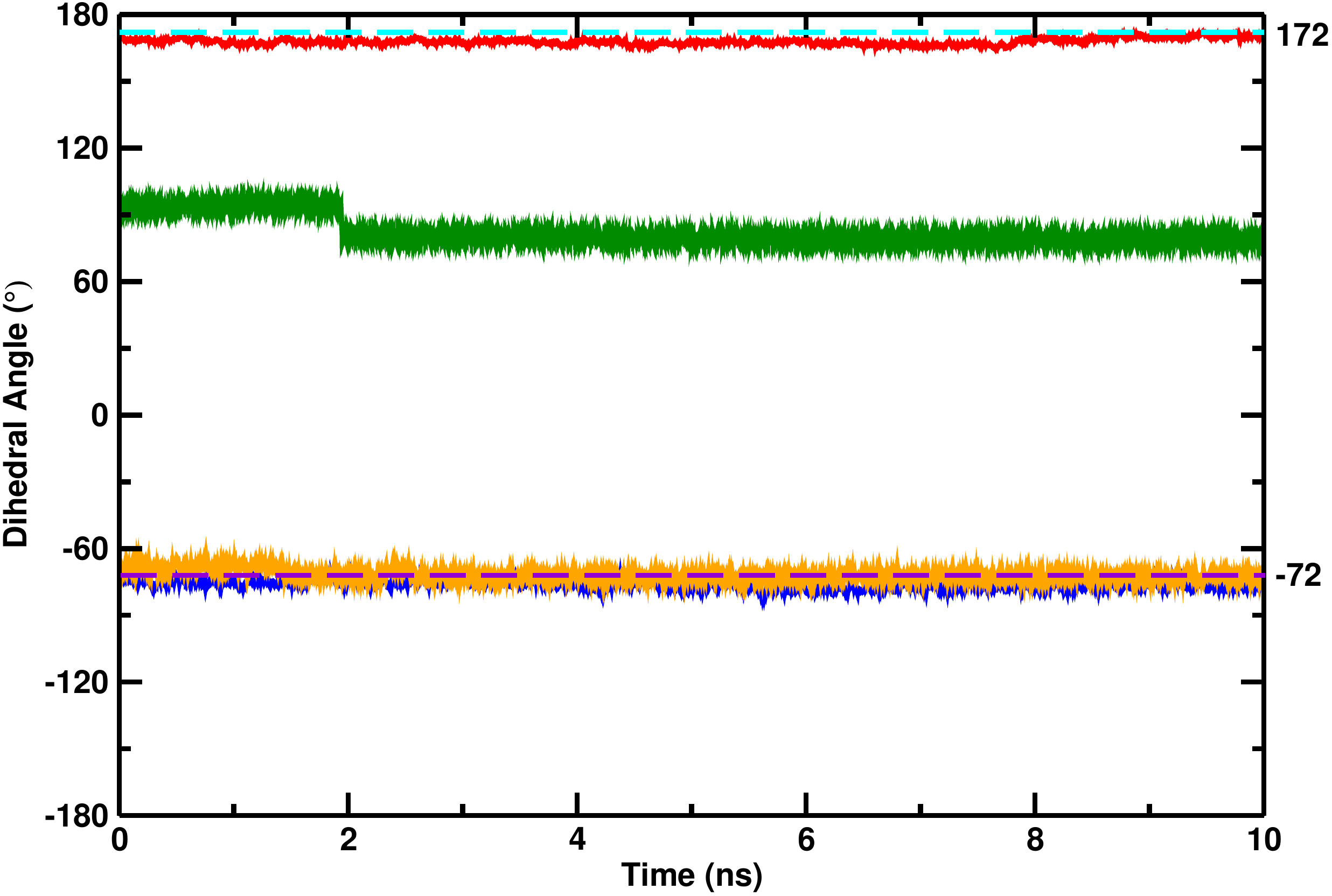}
    \caption{Dihedral angle of $\arcangle$C$_\alpha $C$_\beta$SN for
      Cys10 as a function of time in cis-MbSNO with PC (red),
      trans-MbSNO with PC (blue), cis-MbSNO with MTP (green) and
      trans-MbSNO with MTP (orange) at 50 K. Dashed lines represent
      the values from the major (cyan) and minor (violet) cis-MbSNO
      conformer in 2NRM.}
   \label{sifig:dihedral_50}
\end{figure}

\begin{figure}[H]
\centering
\includegraphics[width=0.99\linewidth]{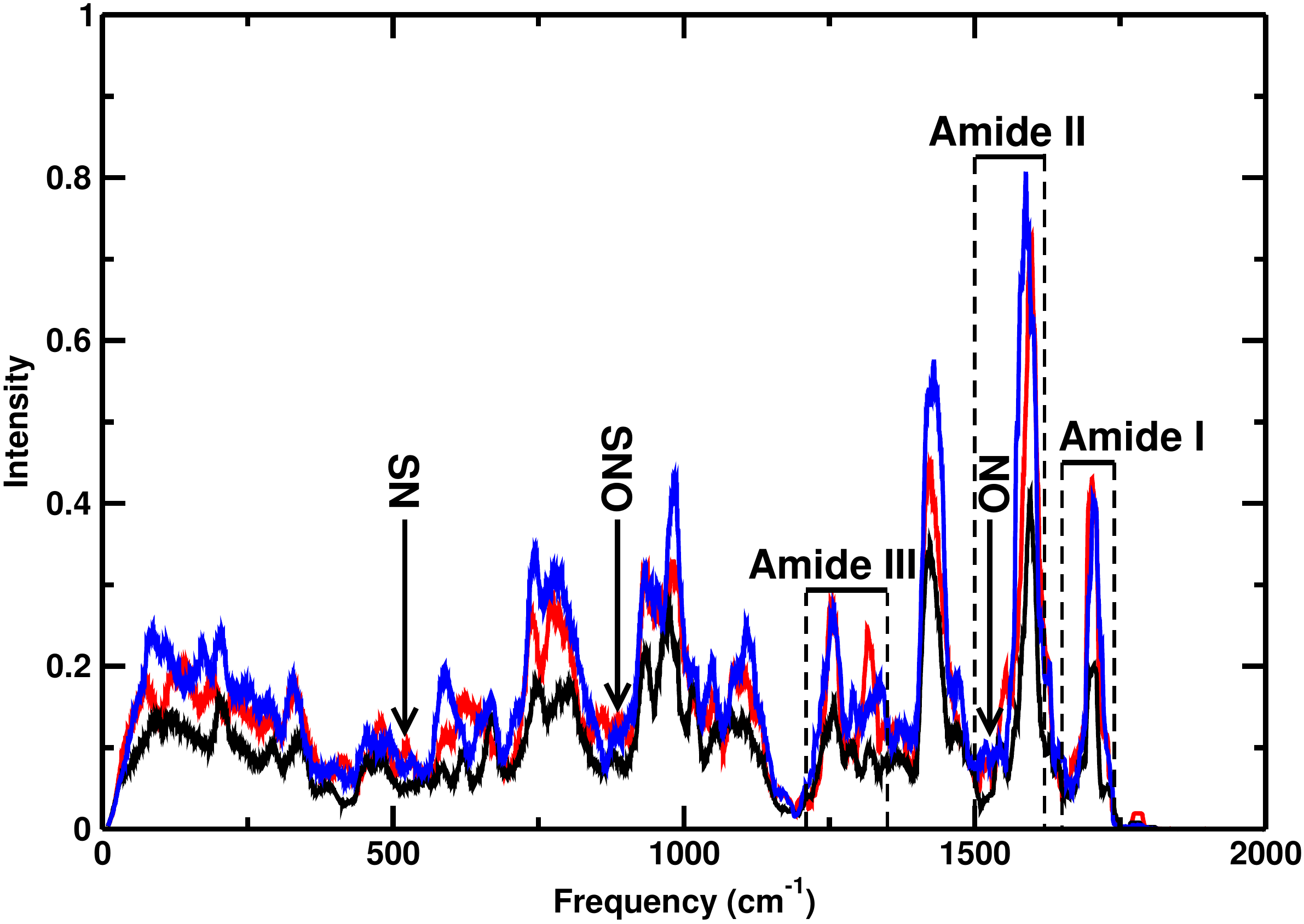}
\caption{Infrared spectrum of WT (black), cis-MbSNO (red) and
  trans-MbSNO (blue) at 50 K from the total dipole moment of the
  protein.}
\label{sifig:mb.ir}
\end{figure}

\begin{figure}[H]
\centering
\includegraphics[width=0.99\linewidth]{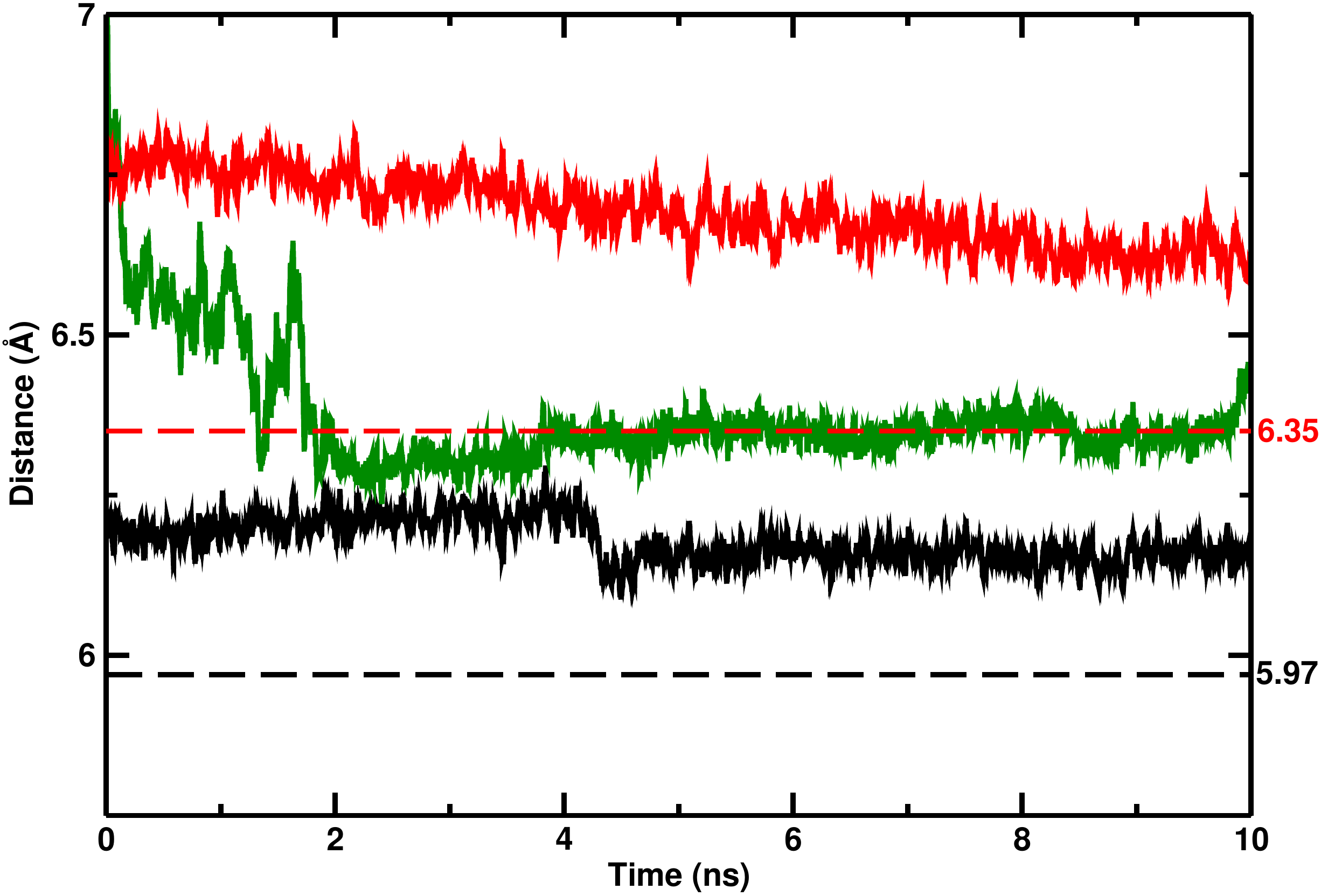}
\caption{Distance between the C$_{\alpha}$ of Cys10 and Ile117
  residues in WT (black), cis-MbSNO with PC (red) and cis-MbSNO with
  MTP (green) at 50 K. Dash lines represents the Cys10 - Ile117
  distance in the crystal structure of WT and major conformer of
  cis-MbSNO in 2NRM}
\label{sifig:ca_dist_50}
\end{figure}

\begin{figure}[H]
\centering
\includegraphics[width=0.99\linewidth]{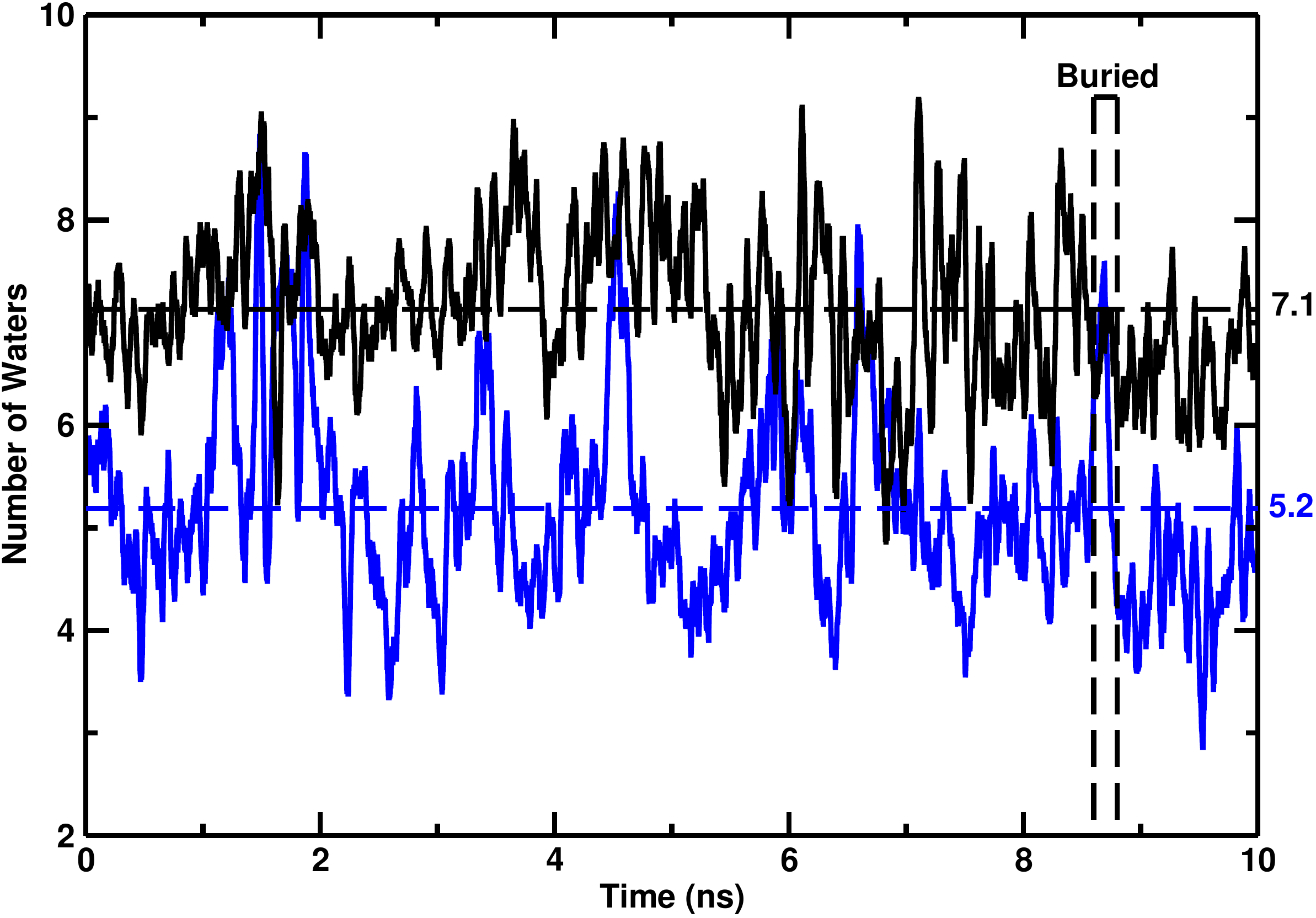}
\caption{The number of water molecules between helices A and H and
  around loop GH. All water oxygen atoms within 5 \AA\/ of residues 5
  to 12 (helix A) and residues 113 to 130 (loop GH, helix H) as a function of
  time are reported for WT (black) and trans-MbSNO (blue, simulation
  with PCs). The average occupation is 7.1 water molecules for WT
  compared with 5.2 for trans-MbSNO, i.e.  a difference of 30 \%.}
\label{sifig:hydration}
\end{figure}